\documentclass[twocolumn,showpacs,amsmath,amssymb]{revtex4}
\usepackage{graphicx}
\usepackage{dcolumn}
\usepackage{bm}

\begin{document} 
\title{Collisionless, phase-mixed, dispersive, Gaussian 
Alfven pulse in transversely inhomogeneous plasma}
\author{D. Tsiklauri}
\affiliation{Astronomy Unit, School of Physics and Astronomy, 
Queen Mary University of London, Mile End Road, London, E1 4NS, United Kingdom}
\date{\today}
\begin{abstract}
   In the previous works 
harmonic, phase-mixed, Alfven wave dynamics was considered
both in the kinetic and magnetohydrodynamic regimes. 
Up today only magnetohydrodynamic, phase-mixed, Gaussian Alfven 
pulses were investigated. 
In the present work we extend this into kinetic regime. 
Here phase-mixed, Gaussian Alfven pulses are studied, which are more appropriate for
solar flares, than harmonic waves,   as the flares are impulsive in nature.
   Collisionless, phase-mixed, dispersive, Gaussian Alfven pulse in transversely 
inhomogeneous plasma is investigated by particle-in-cell (PIC) simulations and 
by an analytical model.
The pulse is in inertial regime with plasma beta less than electron-to-ion 
mass ratio and 
has a spatial width of 12 ion inertial length. 
  The linear analytical model predicts that the pulse
amplitude decrease is described by the linear Korteweg de Vries (KdV) equation. The numerical 
and analytical solution
of the linear KdV equation produces the pulse amplitude decrease in time as $t^{-1}$.
The latter scaling law is corroborated by full PIC simulations. It is shown that the pulse
amplitude decrease is due to dispersive effects,
while electron acceleration is due to Landau damping of the phase-mixed
waves.  
The established amplitude decrease in time as $t^{-1}$ is different from
the MHD scaling of $t^{-3/2}$. This can be attributed to the dispersive effects
resulting in the different scaling compared to MHD, 
where the resistive effects cause the damping, in turn, enhanced by the inhomogeneity.
Reducing
background plasma temperature and increase in ion mass 
yields more efficient particle acceleration.
  \end{abstract}	


\maketitle

\section{Introduction}

Alfven waves are ubiquitous in space and solar plasmas and also,
super-thermal particles play an important role in the same
situations. Some of the examples include:
Earths Auroral zone where
observations show two modes of particle acceleration:
auroral electrons narrowly peaked at specific energy,
implying existence of a static parallel electric field (e.g. \citet{1980SSRv...27..155M});
and observations by FAST spacecraft (e.g. \citet{2007JGRA..11205215C}) 
which show electrons with broad energy and narrow in pitch angle distribution. The latter
suggests that the inertial Alfven wave (IAW) time-varying parallel electric field  accelerates electrons.
Also, in solar corona, about half of the energy released
during solar flares is converted into the energy of accelerated
particles \cite{2004JGRA..10910104E}.
The time-varying parallel electric field maybe produced by 
low frequency ($\omega < \omega_{ci}$, where $\omega_{ci}=eB/m_i$ 
is the ion cyclotron frequency) 
dispersive Alfven waves (DAW) whose wavelength, perpendicular to the background magnetic field,
becomes comparable to any of 
the kinetic spatial scales such as: ion gyro-radius at electron temperature, 
$\rho_s=\sqrt{k_B T_e/m_i}/ \omega_{ci}$, ion thermal 
gyro-radius, $\rho_i=\sqrt{k_B T_i/m_i}/ \omega_{ci}$, 
\cite{1976JGR....81.5083H} or to electron inertial length 
$\lambda_e = c/ \omega_{\mathrm{pe}}$ \cite{1979JGR....84.7239G}. 
Under space plasma nomenclature DAWs are sub-divided 
into Inertial Alfven Waves or Kinetic Alfven Waves (KAW) 
depending on the relation between the plasma $\beta$ and  electron/ion mass 
ratio $m_e/m_i$ \citep{2000SSRv...92..423S}.
When $\beta \ll m_e/m_i$ (i.e. when Alfven speed is much greater 
than electron and ion thermal speeds, 
$C_A \gg v_{th,i}, v_{th,e}$) dominant mechanism for sustaining 
$E_\parallel$ is the parallel electron inertia 
and such waves are called Inertial Alfven Waves. In the opposite case of 
$\beta \gg m_e/m_i$, (i.e. when  $C_A \ll v_{th,i}, v_{th,e}$) 
the thermal effects are more important and the main mechanism for 
supporting  $E_\parallel$ is the parallel electron pressure gradient. Such waves 
are called  Kinetic Alfven Waves.

\citet{2011PhPl...18i2903T} gives an overview of the previous work on this topic
in some detail. \citet{2011JGRA..11600K15M}
studies the interaction of an isolated Alfven wave packet with a plasma density cavity.
\citet{2011PhPl...18i2903T} considered particle acceleration by DAWs in the 
transversely inhomogeneous plasma via full kinetic simulation particularly focusing on
the effect of polarization of the waves and different regimes (inertial and kinetic).
In particular, \citet{2011PhPl...18i2903T}  studied particle 
acceleration by the low frequency ($\omega=0.3\omega_{ci}$)
DAWs, similar to considered in \citet{2005A&A...435.1105T,2008PhPl...15k2902T}, in {\it 2.5D geometry}.
Subsequently,  \citet{2012PhPl...19h2903T} considered 3D effects on 
particle acceleration and parallel electric field generation.
In particular, instead of 1D transverse, to the magnetic field, density (and temperature) inhomogeneity,
the 2D transverse density (and temperature) inhomogeneity was considered. This was in a form of a circular cross-section
cylinder, in which density (and temperature) varies smoothly across the uniform magnetic field
that fills entire simulation domain. Such structure mimics a solar coronal loop which is kept in 
total pressure balance.

As described in \citet{2014PhPl...21e2902T,2011PhPl...18i2903T},
presence and 
damping of magnetohydrodynamic (MHD) waves is of importance to 
several problems:
(i) The solar coronal 
heating problem \cite{2005psci.book.....A},  
(ii) Earth magnetosphere energization
in the context of electron acceleration by Alfven harmonic waves and 
pulses propagating in an auroral plasma cavities
\cite{2011JGRA..11600K15M,refId0,angeo-22-2081-2004,JGRA:JGRA14612}.
(iii) Fast acceleration of inner magnetospheric hydrogen 
and oxygen ions by shock induced ULF waves \cite{2012JGRA..11711206Z}. 
(iv) Heating and stability of Tokamak plasmas, 
e.g. dynamics of 
  shear Alfven waves collectively excited by energetic particles 
  in tokamak plasmas \cite{0029-5515-47-10-S20}.
(v) Heating with waves in the ion cyclotron range of frequencies
is a well-established method on present-day tokamaks and one of the heating
systems foreseen for ITER \cite{PhysRevLett.75.842,
:/content/aip/journal/pop/2/6/10.1063/1.871266,
0029-5515-39-2-306,
2015AIPC.1689c0005M}. 
(vi) It was also 
suggested \cite{2010PPCF...52h5012Q} that off-axis ion 
Bernstein wave heating modifies the electron pressure 
profile and the current density profile can be 
redistributed, 
suppressing the 
magnetohydrodynamic tearing mode instability. 
Such approach provides both the stabilization 
of tearing modes and control of the pressure profiles. 
Phase mixing of harmonic Alfven waves (AW), which propagate in  plasma having a 
density inhomogeneity in  transverse to the uniform background magnetic field direction,
results in their fast damping in the density gradient regions. 
In the  harmonic case the dissipation time scales as 
$\tau_D \propto S^{1/3}$. Where $S= L V/\eta \propto 1/\eta$ is the Lundquist number, 
$\eta=1/(\mu_0 \sigma)$ is plasma resistivity, while
$L$ and $V$ are characteristic length- and velocity- scales of the system. 
This is a consequence of the fact that AW amplitude damps in time as
$B_y(x,z,t)\propto \exp(-\eta C^\prime_A(x)^2 t^3 k^2/6)$, 
where symbols have their usual meaning and $C^\prime_A(x)$ denotes Alfven speed derivative 
in the density inhomogeneity direction
\citep{1983A&A...117..220H}.
Phase mixing of Alfven waves which have Gaussian profile along the background 
magnetic field results in slower, power-law damping,
$B_y\propto t^{-3/2}$, as established by
\citet{2002RSPSA.458.2307W}, and is also derived in more mathematically elegant way in 
 \citet{2003A&A...400.1051T}.

Resuming aforesaid, the motivation for this study is as following:
In the previous works 
harmonic, phase-mixed, Alfven wave dynamics was considered
both in the kinetic 
\citep{2005A&A...435.1105T,2008PhPl...15k2902T,2011PhPl...18i2903T,2012PhPl...19h2903T} and 
magnetohydrodynamic regime \citep{2000A&A...363.1186B}. 
Up today only magnetohydrodynamic, phase-mixed, Gaussian Alfven 
pulses were 
investigated \citep{2002RSPSA.458.2307W,2003A&A...400.1051T,2016A&A...586A..95T}. 
In the present work this is extend into kinetic, dispersive, Alfven pulse regime. 
Thus, phase-mixed, Gaussian Alfven pulses are studied, which are more appropriate for
solar flares, than the harmonic waves,   as the flares are impulsive in their nature.
It is worthwhile noting that
\citet{2011A&A...525A.155T} considered the effect of the Hall term in the 
generalised Ohm's law on the damping and phase mixing of Gaussian Alfven 
pulses in the ion cyclotron range of frequencies in 
uniform and non-uniform equilibrium plasmas. Our work extends the latter
results by considering fully kinetic picture, beyond just the Hall term.
\citet{2009ApJ...693.1494M} explored the possibility that 
electrons could be accelerated by inertial Alfven Gaussian pulses 
to hard X-ray-emitting energies in the low solar corona during flares.
Our work extends the latter reference by including the effect of
transverse inhomogeneity in the Alfven speed, i.e. the effect of phase-mixing.

Section II describes the model for the numerical simulation, 
while the results are presented in section III.
We close with the conclusions in section IV. 

\section{The model}

\begin{figure}[htbp]    
\centerline{\includegraphics[width=0.5\textwidth]{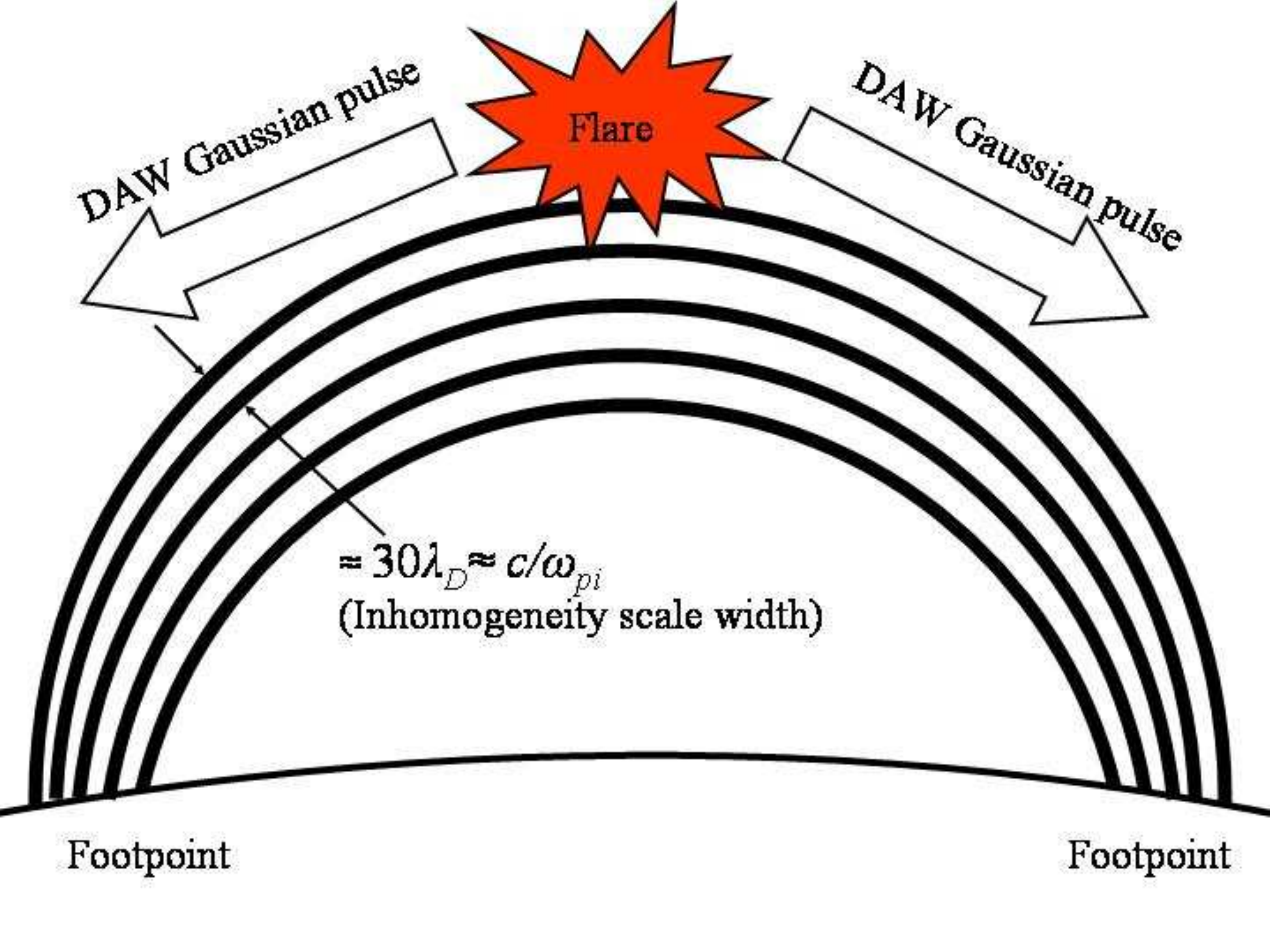}}
\caption{A conceptual sketch of the model. A solar flare launches DAW Gaussian pulses from the 
solar coronal loop apex which rush down towards the photospheric footpoints where X-rays are produced.}
\label{fig1} 
\end{figure}

The general observational context of this work in outlined in Fig.~\ref{fig1}, which shows that a solar flare
at the solar coronal loop apex triggers DAWs Gaussian pulses which then propagate 
towards loop footpoints. 
There are possibilities of
excitation of KAWs or IAWs by means of 
turbulent cascade \cite{2011ApJ...733...33Z} or
magnetic field-aligned currents i.e. essentially electron beams
drifting with respect to stationary ions \cite{2012ApJ...754..123C} or 
fast ion beam excitation \cite{1996SoPh..168..219V}. \citet{2011PhRvE..84d6406C}
considered the situation when KAWs are excited by current (fluid) instability. 
The instability condition for this excitation by current is satisfied 
when the drift velocity, $v_D=0.1v_A$, and KAWs can efficiently grow.
However, \citet{2011PhRvE..84d6406C} 
did not include the resonant excitation of DAWs by the inverse Landau damping 
because its instability condition requires a larger drift velocity, in general, 
larger than the Alfven velocity. 
\citet{2011PhPl...18i2903T,2011A&A...527A.130B} considered a different regime, where
the importance of the Landau (Cerenkov) resonance for
the particle acceleration and parallel electric field generation by the DAWs.

In our model (see Fig.~\ref{fig1}) the transverse density (and temperature) inhomogeneity scale is of the order of
$30$ Debye length ($\lambda_D$) that for the considered mass ratio $m_i/m_e=16$ corresponds to 0.75 
ion inertial length $c/\omega_{pi}$. Possibility of 
existence of such thin loop threads, tens of cm wide, in the solar corona 
is debatable, based on loops observed with TRACE and SDO's AIA.
However, future
high spatial resolution space missions, such as 
Solar Probe Plus and Solar Orbiter may shed light on 
 the possible loop sub-structuring.
Yet another shortcoming, partly associated with previous one,
is the unrealistically small longitudinal scales considered.
Our longest considered  domain is $106.91$ m, which ideally should have
been 100 Mm. Albeit, inability to resolve full kinetics and
realistic spatial scales at the same time,
 plague all current particle-in-cell simulations. 

We use EPOCH (Extendable Open PIC Collaboration) a multi-dimensional, fully electromagnetic, 
relativistic particle-in-cell code which was developed and is used by 
Engineering and Physical Sciences Research Council (EPSRC)-funded 
Collaborative Computational Plasma Physics  (CCPP) consortium of 30 UK researchers \citep{Arber:2015hc}.
We use 2.5D version of the EPOCH code. 
The relativistic equations of motion are solved
for each individual plasma particle.
The code also solves Maxwell's equations, with self-consistent currents, using 
the full component set of EM fields
$E_x,E_y,E_z$  and $B_x,B_y,B_z$.
EPOCH uses SI units. For the graphical
presentation of the results
time is normalised to $\omega_{\mathrm{pe}}^{-1}$.
When visualizing the normalised results we use   $n_0 =10^{16}$ m$^{-3}$ in the
least dense parts of the domain ($y=0$, $y=y_{max}$), which are located at the edges of the
simulation domain (i.e. fix $\omega_{\mathrm{pe}} = 5.64 \times 10^9$ Hz radian on the domain edges).
Here $\omega_{\mathrm{pe}} = \sqrt{n_e e^2/(\varepsilon_0 m_e)}$ is the electron plasma frequency,
$n_\alpha$ is the number density of species $\alpha$ and all other symbols have their usual meaning.
The x-size of the simulation box is different in different numerical runs,
 as stated in table \ref{tab1}.
The considered x-range is
$13200 < x_{max} < 20000$ grid points.  
$y_{max}=200$ grid points and is fixed in all runs. 
The four main runs are for the mass ratio $m_i/m_e=16$.
This mass ratio value corresponds to the in the inertial Alfven wave (IAW) regime
because plasma beta in this study is fixed at 
$\beta= 2 (v_{th,i}/c)^2(\omega_{pi}/\omega_{ci})^2 = 
n_0(0,0)k_B T /(B_0^2/(2\mu_0))=0.02$. Thus $\beta=0.02 < m_e/m_i=1/16=0.0625$;
This is a reasonable compromise value that can be considered with the
available computational resources. 
The grid unit size in the four runs  is $\Delta=\lambda_D$, except for the Run1C where it is 
$\Delta=\sqrt{3}\lambda_D$. 
The latter is because we reduce the background plasma temperature three times and thus 
the Debye length decreases by $\sqrt{3}$. To keep the same length of domain, the factor of 
$\sqrt{3}$ in grid stretching is needed (to keep the same number of grids).
For the four main runs $\lambda_D = v_{th,e}/ \omega_{\mathrm{pe}}=5.345\times 10^{-3}$ m
is the Debye length ($v_{th,e}=\sqrt{k_B T/m_e}$ is electron thermal speed).
This makes the spatial simulation domain size of 
$x=[0,x_{max}]=[0,70.56-106.91m]$, $y=[0,y_{max}]=[0,1.069m]$. 
In the PIC code the velocity of particles
is a continuous physical quantity, however when distribution function
is calculated, this is sampled by a finite velocity (momentum) grid.
Particle velocity space is resolved  
(i.e distribution functions produced in $V_x, V_y, V_z$ directions) with
100000 grid points with particle momenta in the range $\mp 1.5\times10^{-21}$ 
kg m s$^{-1}$ or $\mp 3\times10^{-21}$ kg m s$^{-1}$,
depending on numerical run (see table \ref{tab1}).

We impose constant background magnetic field $B_{0x} =320.0$ Gauss along
$x$-axis. 
This sets $\omega_{ce}/\omega_{\mathrm{pe}} = 0.998$.
Electron and ion temperature at the simulation box edge
is also fixed at $T(0,0)=T_e(0,0)=T_i(0,0)=6\times10^7$K, except for the Run1C where
$T(0,0)=2\times10^7$K. 
This in conjunction with $n_0(0,0) =10^{16}$ m$^{-3}$
makes plasma parameters similar to that of a dense flaring loops
in the solar corona.

\begin{figure}[htbp]    
\centerline{\includegraphics[width=0.5\textwidth]{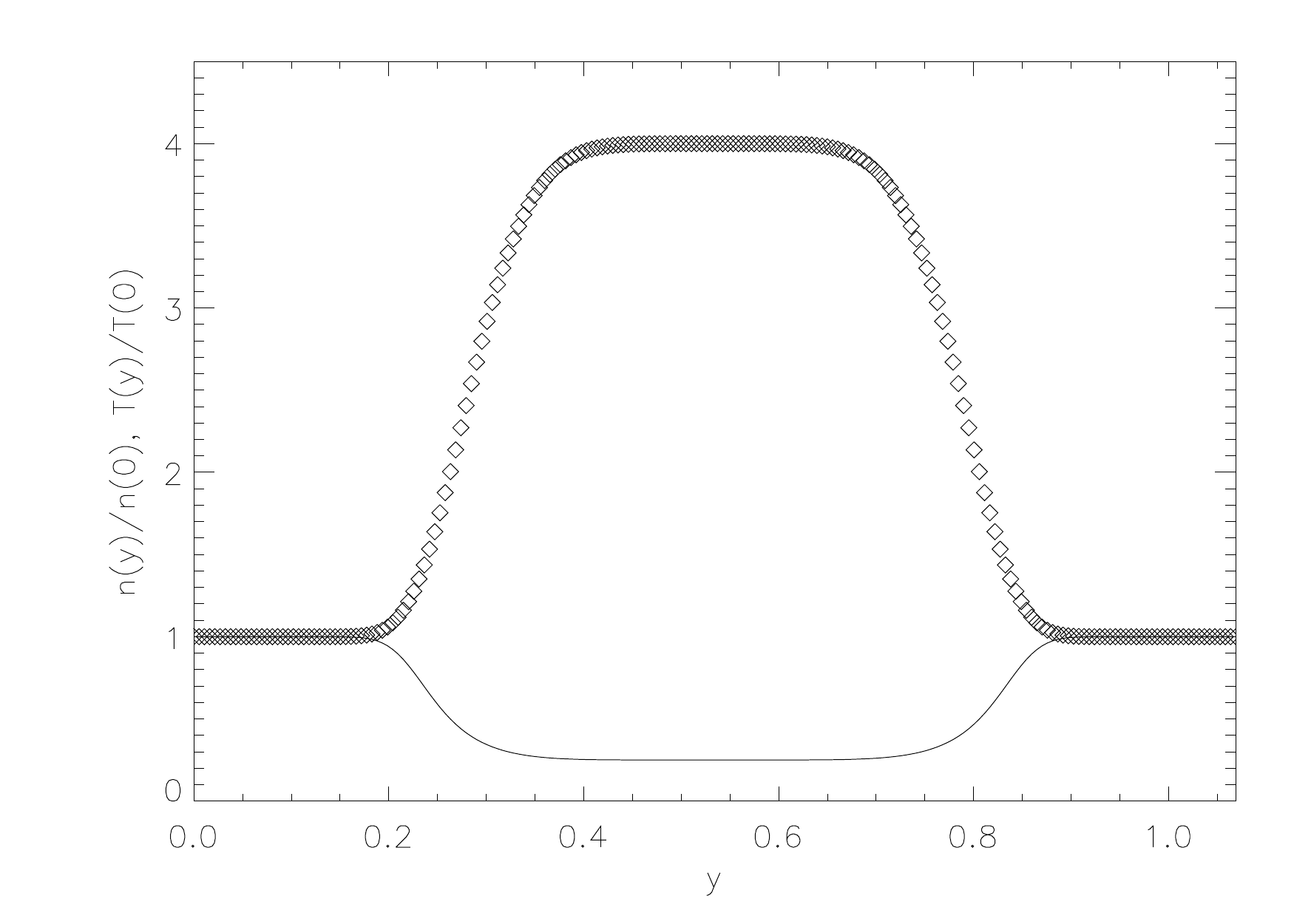}}
\caption{Plot of $n(y)/n(0)$ open diamonds and $T(y)/T(0)$ solid line at $t=0$ according to equations 
\ref{eq1} and \ref{eq2}.}
\label{fig2}
\end{figure}

We consider a transverse to the background magnetic field 
variation of number density as following
\begin{eqnarray}
{n(y)}=
1+3 \exp\Biggl[- \left(\frac{y- y_{max}/2}{50 \Delta}\right)^6\Biggl]
\equiv f(y).&
\label{eq1}
\end{eqnarray}
Equation \ref{eq1} implies that in the central region (across the 
$y$ direction), the density is
smoothly enhanced by a factor of 4, and there are the 
strongest density gradients having 
a width of about ${30 \Delta}$ around the 
points $y=51.5 \Delta$ and $y=148.5 \Delta$, as can be seen in Fig.~\ref{fig2}.

Fig.~\ref{fig2} shows $n(y)/n(0)$ open diamonds and $T(y)/T(0)$ solid line at $t=0$. 
This density behaviour represents the solar coronal loop.
The background temperature of ions and electrons are varied accordingly
\begin{equation}
{T_i(y)}/{T_0}=
{T_e(y)}/{T_0}=f(y)^{-1},
\label{eq2}
\end{equation}
such that the thermal pressure remains constant. 
Because the background magnetic field
along the $x$-coordinate  is constant, the total 
pressure is also constant, ensuring the 
pressure balance. 
Note that flaring solar coronal loops
are not such simple pressure-balanced structures.
In reality during the flare there will be a
magnetic energy release as a result of magnetic
reconnection. It is commonly accepted that 
this will produce heat and super-thermal
particles rushing down towards the sun.
The point often over-looked is that this energy release
results also in magnetic field reconfiguration
launching Alfven waves too \cite{0004-637X-675-2-1645}. Inherent transverse
inhomogeneity will create progressively smaller
spatial scales via phase-mixing \cite{2005A&A...435.1105T}. 
Our simplified initial configuration does not
take into account complex nature of the flare
magnetic energy release. To study wave dynamics
in this idealized pressure-balanced structure
seems a reasonable starting point, but it should be
keep in mind that flaring solar coronal loops
are far from the pressure balance.

The DAW is launched by 
three different means: 
(i) Run 1, with 
driving domain left edge, $x=1\Delta$, with the electric field as follows
\begin{equation}
E_y(1,y,t+\Delta t)= E_y(1,y,t)+ 
E_0 \exp\left(-\frac{(t-(30 /\omega_{ci}))^2}{(12 /\omega_{ci})^2}\right).
\label{eq3}
\end{equation}
Here $E_0=1.4390\times10^6$ V/m that corresponds to 
$0.6 m_e \omega_{\mathrm{pe}} C_A(0,0)/e$ (for $m_i/m_e=16$).
This produces the DAW pulse by electric field driving.
(ii) Run 2, where we impose $B_z$ and $E_y$ Gaussian pulses
\begin{equation}
B_z=0.2 B_0\exp\left(-\frac{(x-(30 c/\omega_{pi}))^2}{(12 c/\omega_{pi})^2}\right), 
\,E_y=B_z C_A(y)
\label{eq4}
\end{equation}
at t=0;
(iii) Run 3, where we impose $B_z$ and $E_y$ Gaussian pulses
as in (ii), plus Alfvenic velocity perturbation $V_z=-C_A(y) B_z$; The latter is essentially achieved
by including additional species of both electrons and ions with particle momentum drifts
of $7.0336\times10^{-23}$ and $1.1254\times10^{-21}$ kg m s$^{-1}$, which correspond to
$p_{e,i}=m_{e,i}C_A(0,0)/\sqrt{1-C_A(0,0)^2/c^2}$. For the mass ratio $m_i/m_e=16$
$C_A(0,0)/c=0.2494\approx0.25$ and for $m_i/m_e=64$
$C_A(0,0)/c=0.1247\approx0.125$. The additional species are localised in x-coordinate
as following
\begin{equation}
n_{e,i}=0.2 n_0
\exp\left(-\frac{(x-(30 c/\omega_{pi}))^2}{(12 c/\omega_{pi})^2}\right).
\label{eq5}
\end{equation}
We never include these additional species in any particle data visualisation, as
only dynamics of background electrons and ions is shown.
As will be shown below when discussing Run 3 such initial conditions achieve nearly perfect
launch of single Gaussian pulse in the positive x-direction. While runs
1 and 2 suffer from the shortcoming that initial Gaussian pulse is split
into two pulses with half-amplitude (positive and negative for run 1 and
both positive for run 2) propagating in the opposite
directions.

\begin{table}[htbp]
\caption{Numerical runs details. 
PPC/TPS stands for particles-per-cell/total-particles-per-species in billions ($10^9$).
$t_{end}$ is simulation end time in units of $\omega_{\mathrm{pe}} $.
$p_{max}$ stands for the considered numerical momentum range $\mp p_{max}$ in units of
$10^{-21}$ kg m s$^{-1}$.} 
\centering 
\begin{tabular}{c c c c c c c} 
\hline\hline 
Case & $n_x $ & PPC/TPS & $t_{end}$  & $p_{max}$ & core$\times$hours & $m_i/m_e$\\ [0.5ex] 
\hline %
Run1   & $13200 $ & $512/1.35$ & $2406$  &1.5 & $192\times 74$  & 16\\ 
Run1H  & $20000 $ & $340/1.36$ & $9622$  &3.0 & $384\times 135$ & 64\\
Run2   & $15000 $ & $512/1.54$ & $2806$  &1.5 & $192\times 110$ & 16\\
Run3   & $15000 $ & $256/0.77$ & $2526$  &1.5 & $192\times 114$ & 16\\ 
Run1C  & $13200 $ & $512/1.35$ & $2406$  &1.5 & $192\times 71$ & 16\\ [1ex]  
\hline
\end{tabular}
\label{tab1} 
\end{table}

\section{Results}

\subsection{theoretical consideration}

As discussed by \citet{2000SSRv...92..423S,2009ApJ...693.1494M} in the inertial regime
($\beta \ll m_e/m_i$) when Alfven perpendicular wavelength approaches the kinetic scales,
electrostatic potential, $\phi$, and magnetic vector potential component along the
background magnetic field $A_x$ satisfy the following equations
\begin{equation}
\left(1-\frac{c^2}{\omega_{\mathrm{pe}}^2} \nabla_\perp^2\right)\frac{\partial A_x}{\partial t}=
-\frac{\partial \phi}{\partial x},
\label{eq6}
\end{equation}
\begin{equation}
\frac{\partial A_x}{\partial x}=
-\frac{1}{C_A(y)^2}\frac{\partial \phi}{\partial t},
\label{eq7}
\end{equation}
where $\nabla_\perp^2=\partial^2_{yy}+\partial^2_{zz}$.
Taking time derivative of equation~\ref{eq6} and then expressing
$\partial^2\phi_{xt}$ from x-derivative of equation~\ref{eq7}, one
arrives at the master equation for $A_x$
\begin{equation}
\left[ \partial^2_{tt} -C_A(y)^2 \partial^2_{xx}\right] A_x
=\left({c^2}/{\omega_{\mathrm{pe}}^2}\right) \partial^2_{yy} \partial^2_{tt} A_x.
\label{eq8}
\end{equation}
Equation~\ref{eq8} left hand side is essentially the wave equation and the right hand side
corresponds to a dispersion. As we will show below the pulse
amplitude decrease is due to dispersive effects. 
It is worthwhile noting that Equation~\ref{eq8}
if formally similar to the resistive MHD case, in particular equation (A.1) from \citet{2003A&A...400.1051T},
with the following substitution $\eta \to (c/\omega_{\mathrm{pe}})^2$ and $\partial_t \to \partial^2_{tt}$. Also
the following relation holds $\vec B=\nabla \times \vec A$.

We now introduce the following coordinates, that are 
co-moving with the wave, as well as
slow (dispersion) time scale:
$(x,y,z,t)\to(\xi, \bar y, \bar z, \tau)$:
$\xi = x-C_A(y)t$,  $\bar y=y$, $\bar z =z$, and $\tau = \varepsilon t$ 
(with $\varepsilon \ll 1$).
The derivatives using the new coordinate system are:
\begin{equation}
\frac{\partial }{\partial x} =\frac{\partial }{\partial \xi}, \,\,\,\,
\frac{\partial }{\partial z} =\frac{\partial }{\partial \bar z},
\label{eq9}
\end{equation}
\begin{equation}
\frac{\partial }{\partial y} = \frac{\partial }{\partial \bar y} -
C_A^\prime(y)t \frac{\partial }{\partial \xi},
\label{eq10}
\end{equation}
\begin{equation}
\frac{\partial }{\partial t} =  - C_A(y)\frac{\partial }{\partial \xi} 
+\varepsilon \frac{\partial }{\partial \tau}. 
\label{eq11}
\end{equation}
Here, prime denotes a derivative over $y$.
Using the co-moving variables and their derivatives, 
the leading term on the left hand side of equation \ref{eq8}
is $-2 \varepsilon C_A(y) \partial^2_{\tau \xi} A_x$.
On the
right hand side we have
\begin{eqnarray}
\frac{c^2}{\omega_{\mathrm{pe}}^2}\left(\frac{\partial^2}{\partial \bar y^2} 
-2C^\prime_A(y)t \frac{\partial^2}{\partial \bar y \partial \xi}+
C^\prime_A(y)^2t^2 \frac{\partial^2}{\partial \xi^2} \right) \times &   \nonumber \\
\left(\varepsilon^2\frac{\partial^2}{\partial \tau^2} 
-2\varepsilon C_A(y)t \frac{\partial^2}{\partial \tau \partial \xi}+
C_A(y)^2 \frac{\partial^2}{\partial \xi^2}\right).& 
\label{eq12}
\end{eqnarray}
Thus keeping the largest terms from the each bracket,
i.e. with $t^2$ in the first (because we consider large times $t/t_{Alfven}\gg 1$) 
and one without $\varepsilon$
in the second bracket,
the leading term is 
$ (c^2/\omega_{\mathrm{pe}}^2) C_A(y)^2 \partial^2_{\xi \xi} [ C^\prime_A(y)^2 
(\tau / \varepsilon)^2 \partial^2_{\xi \xi} A_x]$.
Performing integration over $\xi$ and introduction of yet another
auxiliary variable, $s= (c^2/\omega_{\mathrm{pe}}^2) C^\prime_A(y)^2 C_A(y) \tau^3/ (6 \varepsilon^3)=
(c^2/\omega_{\mathrm{pe}}^2) C^\prime_A(y)^2 C_A(y) t^3/ 6$, we obtain the following 
equation for $A_x$: 
\begin{equation}
\partial_s A_x= - \partial^3_{\xi \xi \xi} A_x.
\label{eq13}
\end{equation}
Equation~\ref{eq13} is the linear Korteweg de Vries (KdV) equation.
Non-linear KdV equation describes propagation of solitons where non-linearity (usual $6 A_x \partial_\xi A_x$ term)
provides wave-overturning, while dispersion (the $\partial^3_{\xi \xi \xi} A_x$ term) causes 
spatial spreading and appearance of wave-forms on the left side of the domain (see figure~\ref{fig7}, 
bottom left panel). It is quite natural that the dispersive term $\partial^3_{\xi \xi \xi} A_x$
appears in the equation describing  dispersive (inertial) Alfven wave.
The relevance here is also in the $c / \omega_{\mathrm{pe}}$ term which replaces the resistivity $\eta$,
compared to the resistive MHD, represents electron  inertial length.
We note that in the resistive MHD phase-mixing the equivalent to equation~\ref{eq13}
is as following
\begin{equation}
\partial_s B_y=  \partial^2_{\xi \xi} B_y,
\label{eq14}
\end{equation}
which is equation (A.3) from \citet{2003A&A...400.1051T}.
Equation~\ref{eq14} is the diffusion equation, because in the
resistive MHD magnetic field diffuses through plasma.
In the homogeneous plasma regions the Alfvenic, Gaussian pulse 
amplitude diffuses as $B_y(t)\propto t^{-1/2}$ \citep{2016A&A...586A..95T}, while due to the
effect of phase mixing, in the inhomogeneous regions the diffusion
is faster, $B_y(t)\propto t^{-3/2}$ \citep{2003A&A...400.1051T}.  

In order to solve liner KdV equation~\ref{eq13} we employ non-unitary Fourier forward and inverse transforms
\begin{eqnarray}
\hat f(k)=\int_{-\infty}^{\infty}f(x)e^{-ikx}dx, &   \nonumber \\
f(x)=\frac{1}{2\pi}\int_{-\infty}^{\infty}\hat f(k)e^{ikx}dk.&
\label{eq15}
\end{eqnarray}
Substituting equation~\ref{eq15} into \ref{eq13} with the initial (at $s=0$ instant) condition,
$A_x(\xi,0)=\exp(-\xi^2)$, we obtain
\begin{eqnarray}
A_x(\xi,s)=\frac{1}{2\pi}\int_{-\infty}^{\infty}\hat A_x(k)e^{ik^3s}e^{ik \xi}dk=&   \nonumber \\
\frac{1}{2\sqrt{\pi}}\int_{-\infty}^{\infty}e^{-k^2/4+ik^3s+ik \xi}dk.&
\label{eq16}
\end{eqnarray}
We used here the fact that Fourier transform of simple Gaussian $\exp(-\xi^2)$
is also a Gaussian in k-space $\sqrt{\pi}\exp(-k^2/4)$. For large
times $t,\tau,s\to \infty$, as done in equation (A.5) from  
\citet{2003A&A...400.1051T}, from the triple sum under the exponent in
equation~\ref{eq16} we keep the largest term $ik^3s$.
Noting that
\begin{equation}
\int_{-\infty}^{\infty} e^{ik^3s} dk=-\frac{2\pi}{\Gamma(-1/3)}s^{-1/3},
\label{eq17}
\end{equation}
and using equation~\ref{eq16}, we obtain asymptotic (the large times) solution for $A_x$ as following
\begin{equation}
A_x(\xi,s)=-\frac{\sqrt{\pi}}{\Gamma(-1/3)}s^{-1/3}.
\label{eq18}
\end{equation}
Here $\Gamma(-1/3)=-4.06235$ is the Gamma-function.
Thus the main conclusion of this sub-section is that as equation~\ref{eq18}
asserts, Gaussian pulse amplitude scales in time as $\propto s^{-1/3} \propto t^{-1}$.

\subsection{numerical validation}

\begin{figure*}[htbp]    
\centerline{\includegraphics[width=0.75\textwidth]{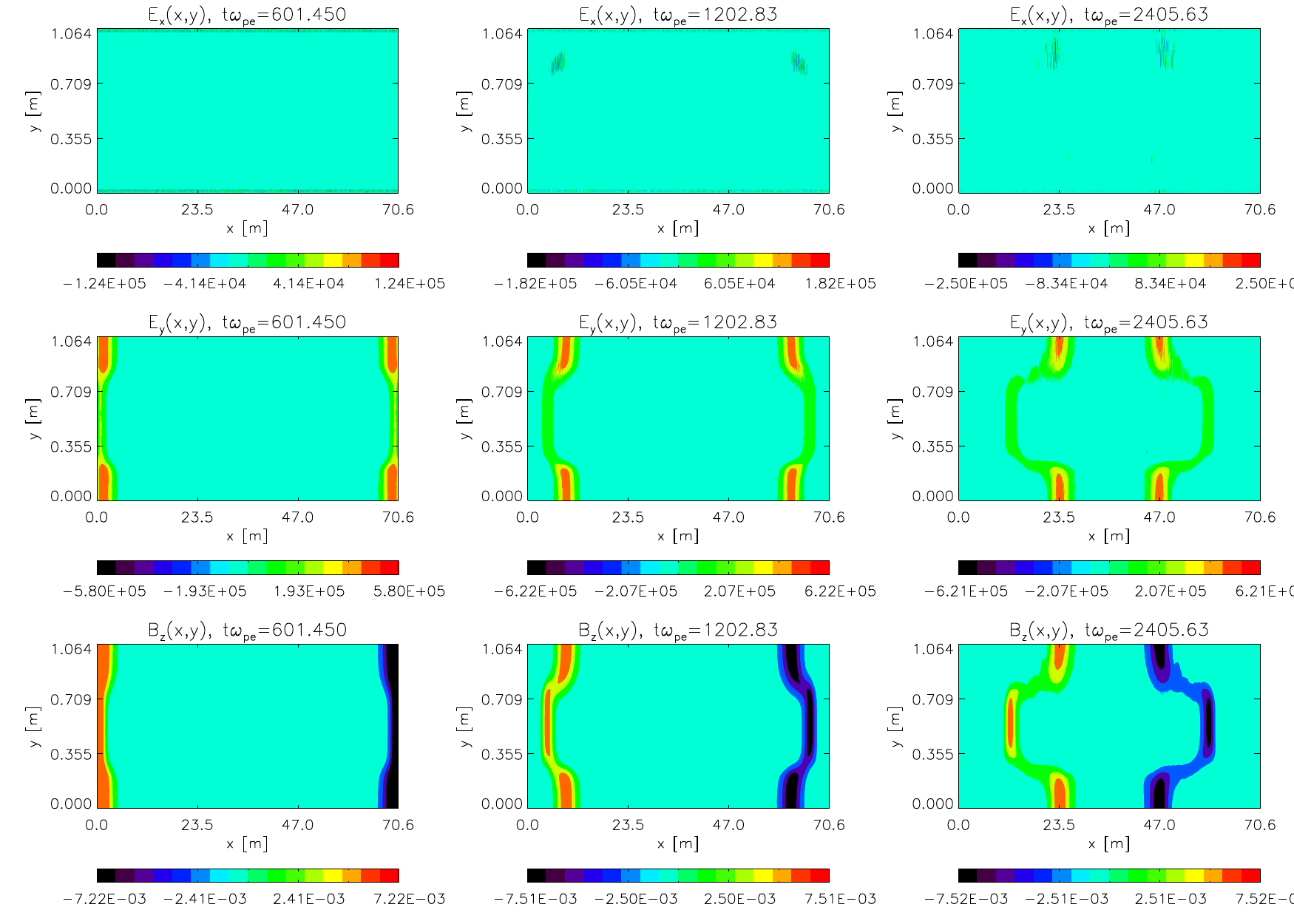}}
\caption{Contour (intensity) plots of the following physical quantities at 
times indicated at individual panels:
(top row) $E_x(x,y)$,
(middle row) $E_y(x,y)$, and
(bottom row) $B_z(x,y)$.}
\label{fig3}
\end{figure*}

Fig.~\ref{fig3} presents the numerical simulation results for the electromagnetic fields for Run 1. 
In Run 1 we output 20, equally spaced in time data snapshots 
and each column in the figure corresponds to 5th, 10th, 20th snapshot.
The 20th snapshot corresponds to $t=2406 \omega_{\mathrm{pe}}=150 \omega_{ci}$. The other parameters of this
run are indicated in Table~\ref{tab1}.
We see from
Fig.~\ref{fig3} the usual phase mixing picture, i.e. $E_y$ and $B_z$ pulses are
excited that propagate in positive and negative x-direction, because the
electric field driving according to equation~\ref{eq3} is unable to excite
a clear eigen-mode of the system propagating on in the positive
x-direction. As throughout the paper we use periodic boundary
conditions, the pulse that propagates to the left, i.e. in the
negative x-direction, re-appears on the right side of the domain and
propagates to the left. The pulse that propagates to the right (positive x-direction)
is also clearly present. Because according to equation~\ref{eq1} the density is
smoothly enhanced by a factor of 4 in the middle of y-coordinate range,
phase speed of the wave is slower there,  as roughly $V_{\rm phase}\approx B_0/\sqrt{\mu_0 m_i n_0}=C_A(y)$.
Thus the front phase-mixes and creates transverse gradients.
It is in the region of these gradients, near  the 
points $y=51.5 \Delta$ and $y=148.5 \Delta$, $E_x$, the parallel electric
field is generated. As seen in the middle and right panels of top row of 
Fig.~\ref{fig3}, the generated $E_x$ is clearly seen near $y=148.5 \Delta$
(but not near $y=51.5 \Delta$ -- this is probably due to small number of contour levels used
to reduce the figure disk space). The similar type behaviour, but for the {\it harmonic} Alfvenic wave
was seen in \citet{2005A&A...435.1105T,2008PhPl...15k2902T}, in 2.5D geometry and 3D
geometry in \citet{2012PhPl...19h2903T}.

\begin{figure}[htbp]    
\centerline{\includegraphics[width=0.5\textwidth]{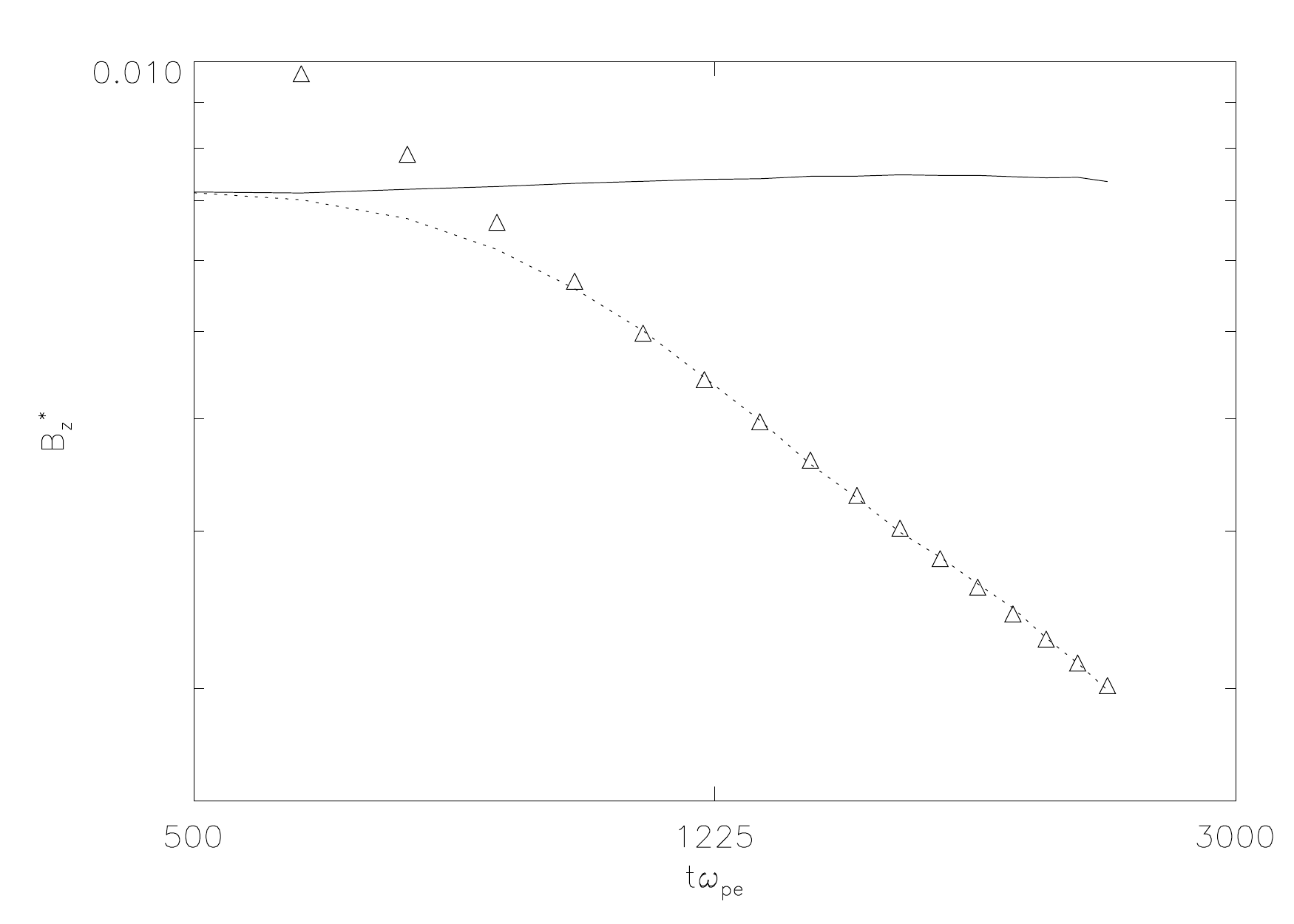}}
\caption{Scaling of $B_z$ DAW pulse amplitude with time. The dashed line
is $B_z^*=\max (B_z(x,y=51 \Delta))$, which tracks the amplitude
in the strongest density gradient point. Solid line is
$\max (B_z(x,y=1 \Delta))$ which tracks the amplitude
away from the density gradient. The triangles show
the analytical (numerically fitted) scaling law 
$B_z^*\propto t^{-1.1}$.}
\label{fig4}
\end{figure}

Fig.~\ref{fig4} shows the scaling of $B_z$ DAW pulse amplitude with time. The dashed line
is $B_z^*=\max (B_z(x,y=51 \Delta))$ for the right propagating pulse, which tracks the amplitude
in the strongest density gradient point. Solid line is
$\max (B_z(x,y=1 \Delta))$ which tracks the amplitude
away from the density gradient. This shows no decrease
(the solid line stays at the same level) meaning that there is no significant amplitude
decrease of
the pulse away from the transverse density gradient regions.
The triangles show
the analytical (numerically fitted) scaling law 
$B_z^*\propto t^{-1.1}$. The fitting is done
using IDL's 
\verb;poly_fit; routine, and 
employing the last 10
out of 20 total time sampling points.
The  \verb;poly_fit; routine
is used as following:
 $B_z=e^r_0 t^{r_1}$ where $r_0$ and $r_1$
the fit parameters. Further we use $\ln B_z=r_0+r_1 \ln t$.
Then using the first order polynomial fit of the form $f(x)=a_0+a_1 x$ in \verb;poly_fit; routine
where $e^r_0=a_0$ and $r_1=a_1$ provides the fit.
Thus we show  that in the kinetic
regime the scaling law for the Gaussian pulse amplitude decay in time 
is not the same as in MHD ($B_z\propto t^{-3/2}$ 
\citet{2016A&A...586A..95T}), namely, $B_z\propto t^{-1}$. This is due to the fact
that the diffusion equation is replaced by the linear KdV equation.
It is worthwhile noting that performing similar fit
to the other strongest gradient point 
$B_z^*=\max (B_z(x,y=148 \Delta))$, not shown here,
produces the best fit of $B_z^*\propto t^{-1.0}$. Thus the results for
both strongest gradient points are consistent.

\begin{figure*}[htbp]    
\centerline{\includegraphics[width=0.75\textwidth]{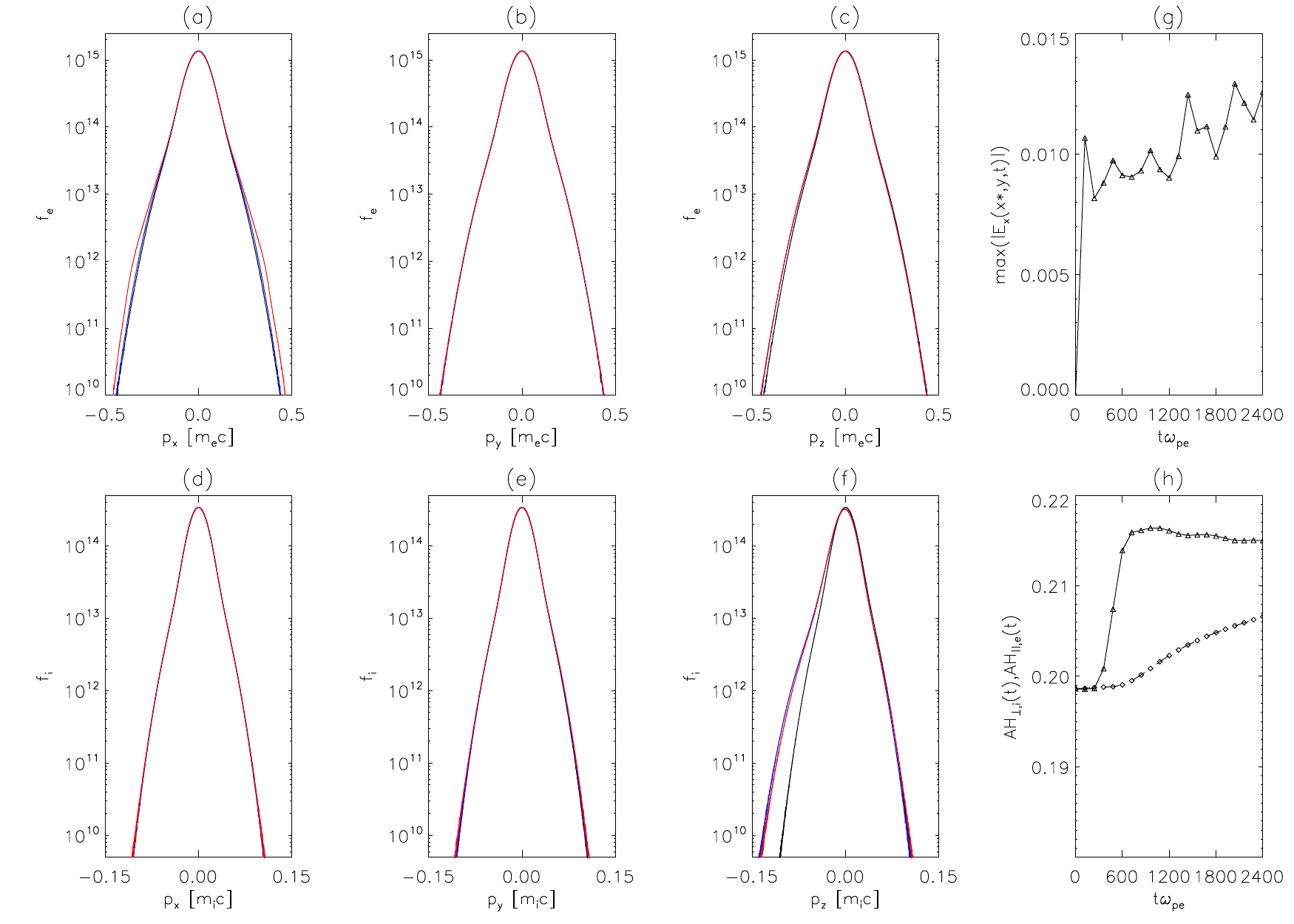}}
\caption{Time evolution of electron and ion velocity distribution 
functions versus velocity $x,y$ and $z$ components on a log-linear plot: 
(a) $f_e(p_x,t=0)$ black (inner) curve, $f_e(p_x,t=t_{end}/2)$ blue  and $f_e(p_x,t=t_{end})$ red (outer) curve,
(b) $f_e(p_y,t=0)$ black curve, $f_e(p_y,t=t_{end}/2)$ blue  and $f_e(p_y,t=t_{end})$ red curve,
(c) $f_e(p_z,t=0)$ black curve, $f_e(p_z,t=t_{end}/2)$ blue  and $f_e(p_z,t=t_{end})$ red curve, 
(d) $f_i(p_x,t=0)$ black curve, $f_i(p_x,t=t_{end}/2)$ blue  and $f_i(p_x,t=t_{end})$ red curve,
(e) $f_i(p_y,t=0)$ black curve, $f_i(p_y,t=t_{end}/2)$ blue  and $f_i(p_y,t=t_{end})$ red curve,
(f) $f_i(p_z,t=0)$ black curve, $f_i(p_z,t=t_{end}/2)$ blue  and $f_i(p_z,t=t_{end})$ red curve.
Time evolution (at 20 time intervals between $t=0$ and $t=t_{end}$) of the following:
(g) $\max(|E_x(x=51 \Delta, y, t|)$, triangles connected with a solid curve
and (h) $AH_{||,e}$ index, diamonds connected with
dashed curve, according to equation~\ref{eq19}, $AH_{\perp,i}$
index, triangles connected with a solid curve, according to equation~\ref{eq20}.
This figure pertains to the numerical Run 1.}
\label{fig5}
\end{figure*}

Fig.~\ref{fig5} shows electron (panels a--c) and ion (panels d--f) distribution function dynamics for Run 1. 
It also shows, in panel (g), the time evolution (at 20 time intervals
between $t=0$ and $t=t_{end}$) of 
$max(|E_x(x=51\Delta, y, t|)$ plotted with  triangles connected with a solid curve.
In panel (h) $AH_{||,e}$ index is plotted with diamonds connected with
dashed curve, according to equation~\ref{eq19}, and $AH_{\perp,i}$ index plotted with triangles
connected with a solid curve, according to equation~\ref{eq20}.
We deduce from panel (a)
that the two bumps in the parallel electron distribution function (for negative and
positive velocities) can be understood by a Landau resonance
because it corresponds to phase speed of the DAW $V_{phase}=0.2494 c$. This is similar 
result to 
to our earlier
works \cite{2005A&A...435.1105T,2008PhPl...15k2902T,2011PhPl...18i2903T,2012PhPl...19h2903T}
for the harmonic Alfven wave. However, what is different (e.g. compare figure~\ref{fig5} to figure~3 from 
\citet{2011PhPl...18i2903T}) is that electric field is twice as weak (panel (g))
and particle acceleration is much less efficient. 
A more rigorous proof that indeed we deal with the Landau
resonance is presented below when discussing Run1H.
We quantify the  particle acceleration by introducing the following quantities:
\begin{equation}
AH_{\parallel,e}(t)=
\frac{\int_{|v_{x}| > \langle v_{th,e}\rangle}^\infty      f_e(v_x,t)dv_x}
 { \int_{-\infty}^\infty   f_e(v_x,0) dv_x},
\label{eq19}
\end{equation}
\begin{equation}
AH_{\perp,i}(t)=
\frac{\int_{|v_{\perp}| > \langle v_{th,i}\rangle}^\infty      f_i(v_\perp,t)dv_\perp}
 { \int_{-\infty}^\infty   f_i(v_\perp,0) dv_\perp },
\label{eq20}
\end{equation}
where $f_{e,i}$ are electron or ion velocity distribution functions and
$<>$ brackets denote average over y-coordinate, because temperature and density vary across y-coordinate.
These definitions 
effectively provides the fraction
(the percentage) of super-thermal electrons and ions.
We gather from panel (h) that 
 $AH_{||,e}$ index  starts from $0.199$ and stops at $0.207$, meaning that 
 the difference $0.207-0.199=0.008\approx 0.01$, i.e.
one percent of electrons are accelerated above thermal speeds.
For ions this number is about twice as large ($\approx 2\%$) 
due to negative $p_z$ momenta in panel (f).

\begin{figure}[htbp]    
\centerline{\includegraphics[width=0.5\textwidth]{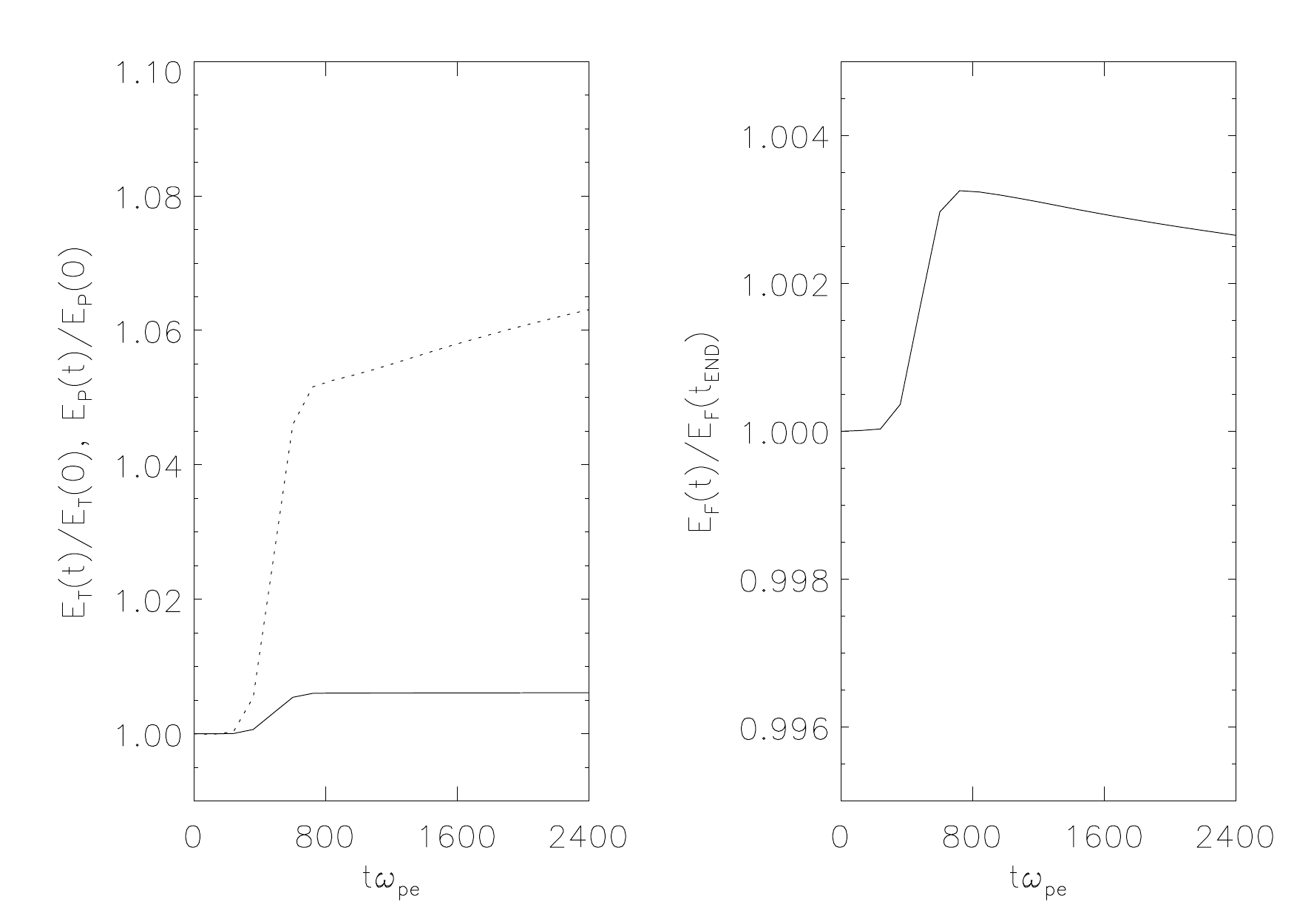}}
\caption{Left panel's solid and dashed curves
are the total (particles plus EM fields) and 
particle energies, normalized on initial values, respectively. 
Right panel shows
EM field energy normalized on its initial value.}
\label{fig6}
\end{figure}

Fig.~\ref{fig6} shows that the total energy has a small, 0.6\%, increase. 
This is a tolerable error due to the well-know numerical heating inherent to PIC codes. 
The all three  energies go up until about $t=800 \omega_{\mathrm{pe}}$
this corresponds to the timescale of growth of the electric field
according to equation~\ref{eq3}. Then particle energy
continues to grow much slowly while particles are accelerated
via collisionless 
Landau damping. The electromagnetic energy decreases after $t=800 \omega_{\mathrm{pe}}$
 which means that 
particle acceleration is on the expense of electromagnetic energy decrease.

\begin{figure*}[htbp]    
\centerline{\includegraphics[width=0.75\textwidth]{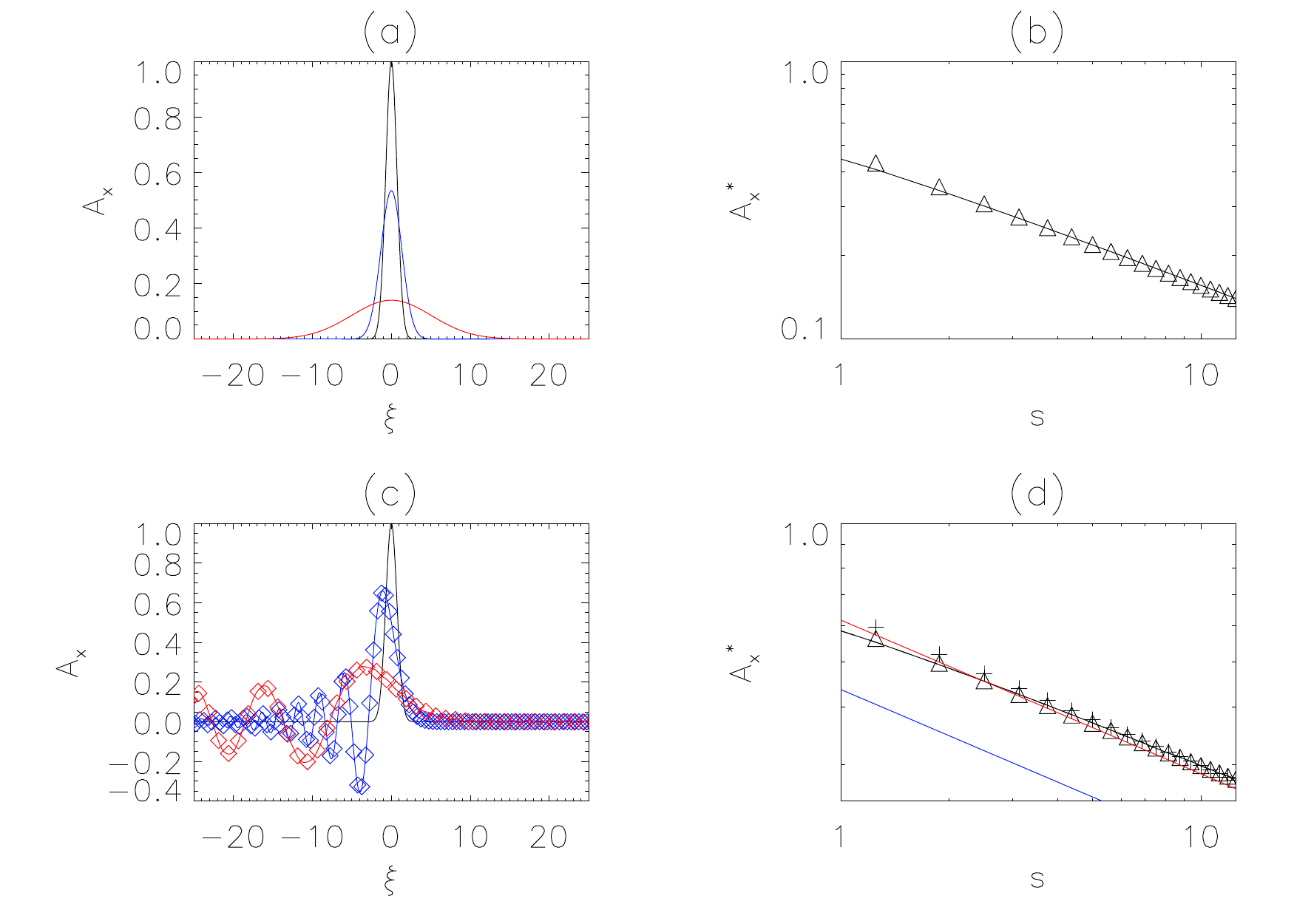}}
\caption{Time evolution of Gaussian pulse, $A_x(\xi,0)=\exp(-\xi^2)$: panel (a) 
according to the diffusion equation~\ref{eq14}.
Black curve is for $A_x(\xi,s=0)$, blue for $A_x(\xi,s=0.625)$ and
red for $A_x(\xi,s=12.5)$. Panel (b) is a log-log plot of $A_x^*=\max(A_x(\xi,s))$
at different times $s$, showing evolution according to the
diffusion equation~\ref{eq14} (with $B_y$ replaced by $A_x$). 
The solid line is numerical solution of the code and
triangles represent numerical fit $s^{-0.488}$ performed
using IDL's poly-fit routine (see text for details).
 Panel (c) shows evolution of $A_x$ according to 
the KdV equation~\ref{eq13}. In addition to
black, blue and red lines, which represent the numerical code solution,
we also plot analytical
solution with open diamonds according to real part of
equation~\ref{eq16}, i.e.
$A_x(\xi,s)=\frac{1}{2\sqrt{\pi}}\int_{-\infty}^{\infty}e^{-k^2/4}\cos(k^3s+k \xi)dk$
using IDL's int-tabulated routine.
 Panel (d) is showing a log-log plot of $A_x^*=\max(A_x(\xi,s))$ according to 
the KdV equation~\ref{eq13}.
The solid black line corresponds to numerical 
solution of the code. In addition, dashed line that fully overlaps
the solid line represents the same analytical solution using IDL's
int-tabulated routine.
The triangles show  represent numerical fit $s^{-0.305}$,
using IDL's poly-fit routine to the
numerical code solution.
The crosses represent the similar fit but to the 
analytical solution and the fit yields $s^{-0.332}$. Blue line is according 
to equation~\ref{eq18} with $s^{-1/3}$. The red line is the same but with 
additional factor $\sqrt{2}$ which fits data better.}
\label{fig7}
\end{figure*}

Fig.~\ref{fig7} presents time evolution of Gaussian pulse, 
$A_x(\xi,0)=\exp(-\xi^2)$. The 
panel (a) is the evolution according to the
diffusion equation~\ref{eq14} (with $B_y$ replaced by $A_x$)
for the numerical code versification purposes.
Black curve is for $A_x(\xi,s=0)$, 
blue for $A_x(\xi,s=0.625)$ and
red for $A_x(\xi,s=12.5)$. Note how the diffusive spatial spread 
(widening of the pulse) progressively becomes evident.
The simple numerical code and Interactive Data Language (IDL) routines are
available from 
\url{http://ph.qmul.ac.uk/~tsiklauri/toyama_2016}.
The code solves the diffusion and KdV equations
using 4th order Runge-Kutta time step.
It uses 4th order centered differencing for the
second order spatial derivative
in the diffusion equation and
2nd order centered differencing for the
third order spatial derivative
in the KdV equation using Table 1 from \citet{fornberg}.
The code has domain size of $\xi=[-250,250]$ and end simulation
times is $t=12.5$. The spatial domain has 20000 points.
Panel (b) is a log-log plot of $A_x^*=\max(A_x(\xi,s))$
at different times $s$,  showing evolution according to the
diffusion equation~\ref{eq14} (with $B_y$ replaced by $A_x$). The  
solid line is numerical solution of a code and
triangles represent numerical fit $s^{-0.488}$,
using IDL's 
\verb;poly_fit; routine, and 
employing the last 6
out of 20 total time sampling points.
In the homogeneous plasma regions the Alfvenic, Gaussian pulse 
amplitude diffuses as $B_y(t)\propto t^{-1/2}$ \cite{2016A&A...586A..95T}.
We see that solution of the diffusion equation is handled very well by the code,
as $-0.488$ is nearly $-0.5$.
We use the last 6 points because the predicted scaling becomes
progressively better for increasing time.
 Panel (c) shows evolution of $A_x$ according to 
the KdV equation~\ref{eq13}. In addition to
black, blue and red lines, which represent the numerical code solution, 
we also plot analytical
solution with open diamonds according to real part of
equation~\ref{eq16}, i.e.
$A_x(\xi,s)=\frac{1}{2\sqrt{\pi}}\int_{-\infty}^{\infty}e^{-k^2/4}\cos(k^3s+k \xi)dk$ using IDL's
\verb;int_tabulated; routine.
The routine uses domain size of $k=[-250,250]$ with 40000 point discretization.
We see that the pulse amplitude decreases but also
the dispersion creates wave-like pattern in the negative $\xi$
region. This is the expected behaviour and we see similar pattern in
PIC simulations too. The diamonds are plotted with much
less than actual grid number points to aid the
visualization (we plot every 20th point for blue diamonds
and every 50th for the red). To a plotting precision
the match between the numerical and analytical solutions
is obvious.
Panel (d) shows a log-log plot of $A_x^*=\max(A_x(\xi,s))$, according to 
the KdV equation~\ref{eq13}.
The solid black line corresponds to numerical 
solution of the code.
In addition, dashed line that fully overlaps
the solid line represents the analytical solution
$A_x(\xi,s)= \frac{1}{2\sqrt{\pi}}\int_{-\infty}^{\infty}e^{-k^2/4}\cos(k^3s+k \xi)dk$ using IDL's
\verb;int_tabulated; routine.
The 
triangles represent numerical fit $s^{-0.305}$,
using IDL's \verb;poly_fit; routine to the
numerical code solution, and 
employing the last 6
out of 20 total time sampling points.
Crosses represent the similar fit but now applied to the 
analytical solution. The fit yields
$s^{-0.332}$. Again, the  \verb;poly_fit; routine
is used in the following manner:
We start from $A_x=e^r_0 s^{r_1}$ where $r_0$ and $r_1$
the fit parameters. Then $\ln A_x=r_0+r_1 \ln s$.
Using first order polynomial fit of the form $f(x)=a_0+a_1 x$ in \verb;poly_fit; routine
where $e^r_0=a_0$ and $r_1=a_1$ gives the desired fitting.
Blue line shows the solution according to
equation~\ref{eq18}. The red line is the solution according to
equation~\ref{eq18} but with additional factor $\sqrt{2}$
which fits data better. Thus we conclude that as equation~\ref{eq18}
shows, the Gaussian pulse amplitude scales in time according to KdV equation~\ref{eq13} as 
$\propto s^{-1/3} \propto s^{-0.3333} \propto t^{-1}$.
While  the homogeneous diffusion equation solution scales as $\propto s^{-1/2}$.

\begin{figure*}[htbp]    
\centerline{\includegraphics[width=0.75\textwidth]{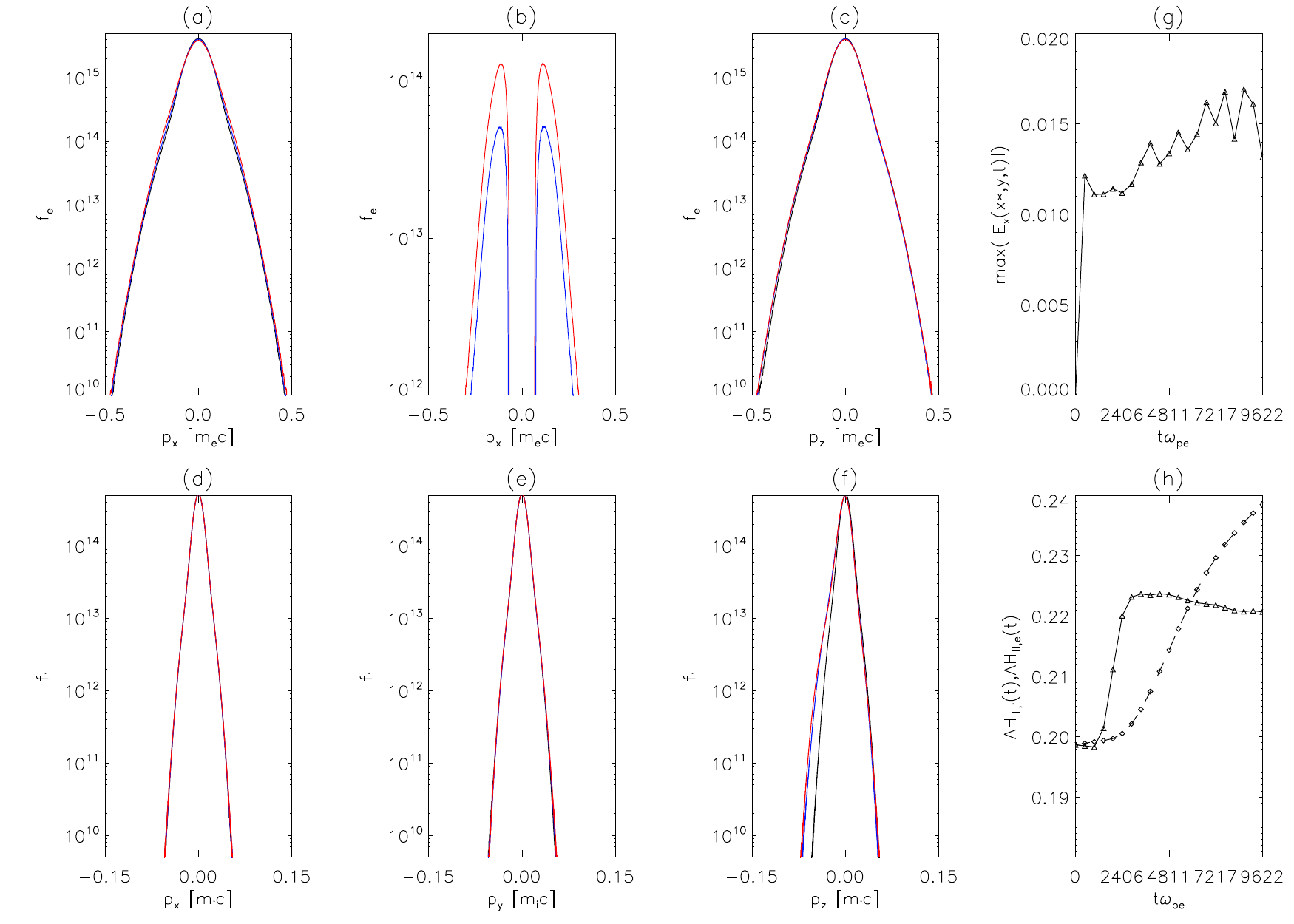}}
\caption{As in Fig.~\ref{fig5}, but for the Run1H with mass ratio 64.
Here we replaced panel (b) with
$f_e(p_x,t=t_{end}/2)-f_e(p_x,t=0)$ blue  and $f_e(p_x,t=t_{end})-f_e(p_x,0)$ red curves, 
in oder to emphasize the difference between the distribution functions
at different times.}
\label{fig8}
\end{figure*}

The Run1H corresponds to keeping everything the same is in Run 1 but
making heavier ions by factor of 4, i.e. now the mass ratio
is $m_i=m_e=64$. This makes the phase speed of DAW twice as small
$V_{phase}/c\approx0.125$.
Fig.~\ref{fig8} is similar to Fig.~\ref{fig5}, but for the Run1H.
It is barely visible that the two bumps in the parallel to the field distribution function, $f_e(p_x)$,
have now shifted from 0.25 to 0.125. Thus we replaced panel (b) with
$f_e(p_x,t=t_{end}/2)-f_e(p_x,t=0)$ blue  and $f_e(p_x,t=t_{end})-f_e(p_x,0)$ red curve, 
in oder to stress the difference between the distribution functions
at different times. It is evident the the bumps are now near $\pm 0.125$. This proves that
the acceleration of the particles is due to Landau damping. Similar conclusion 
also was reached before, when the harmonic DAW was considered 
in \citet{2005A&A...435.1105T,2008PhPl...15k2902T}, in 2.5D geometry and 3D
geometry in  \citet{2012PhPl...19h2903T}.
Other noteworthy feature is the for Run1H we see
more efficient particle acceleration, as 
$AH_{||,e}$ index  starts from $0.199$ and stops at $0.238$, meaning that $0.238-0.199= 0.039\approx0.04$, i.e.
four percent of electrons are accelerated above thermal speeds.
Thus four times more massive ions result in four times more efficient
electron acceleration. Of course, this is still far
below of requirement to produce X-rays in solar flares, where $\approx 50$
percent of electrons are accelerated. It is unclear what the results would
be for the realistic mass ratio of $m_i/m_e=1836$.

\begin{figure*}[htbp]    
\centerline{\includegraphics[width=0.75\textwidth]{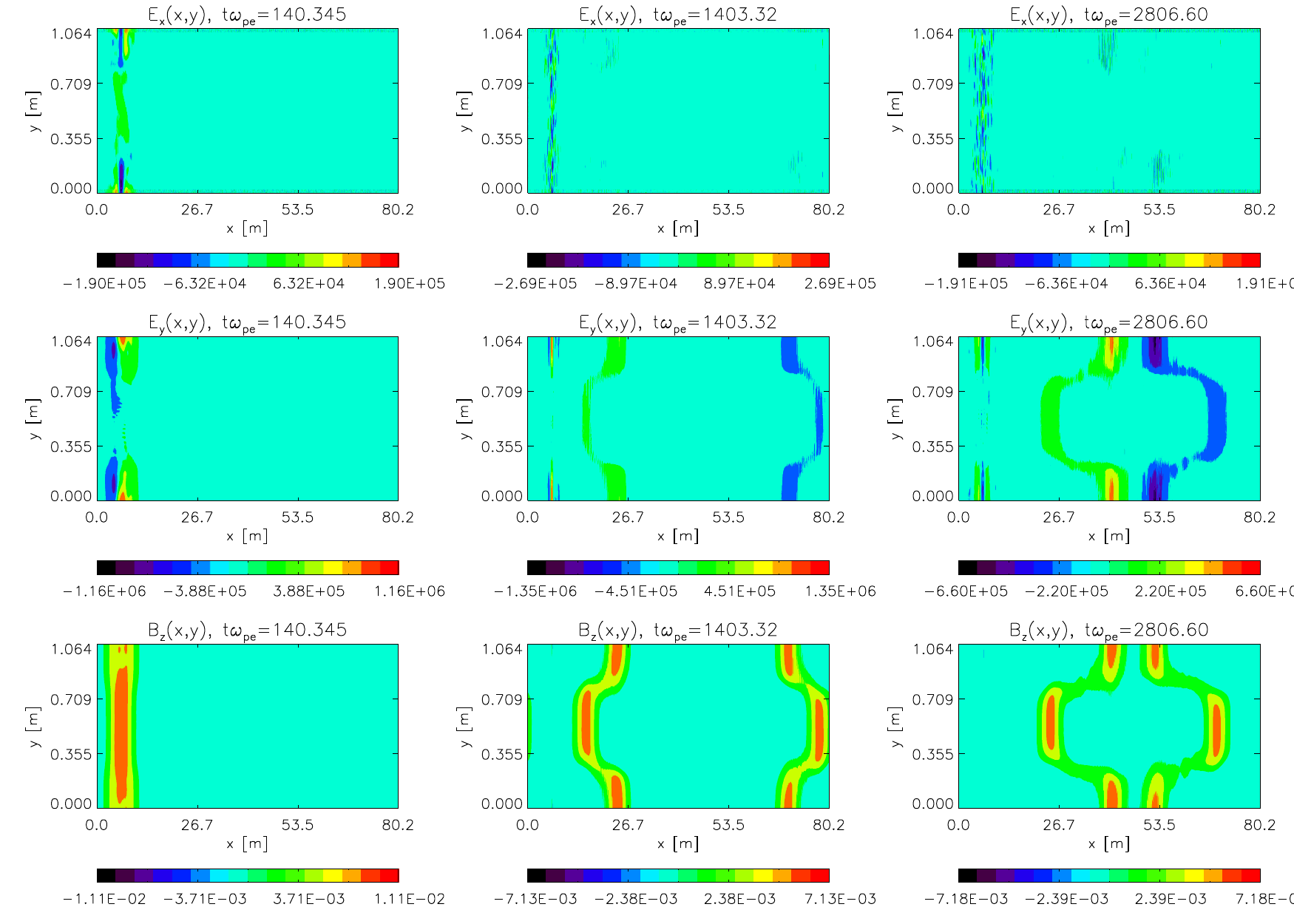}}
\caption{As in Fig.~\ref{fig3} but for the run 2.}
\label{fig9}
\end{figure*}
Fig.~\ref{fig9} presents the numerical simulation results for the electromagnetic fields for Run 2. 
In Run 2 we also output 20, equally spaced in time data snapshots 
and each column in the figure corresponds to 1st, 10th, 20th snapshot.
$E_y$ and $B_z$ show again phase-mixed behaviour, but the new type of
initial conditions according to equation~\ref{eq4} now only result
in a single $B_z$ positive pulse being split into two positive pulses with
half amplitude travelling in opposite directions. In the case of Run 1, the pulse
the moved to the right was positive while one moving to the left was negative.
Parallel electric field is also generated in the transverse 
density gradient regions and now seen about around  
$y=51.5 \Delta$ and $y=148.5 \Delta$.

\begin{figure}[htbp]    
\centerline{\includegraphics[width=0.5\textwidth]{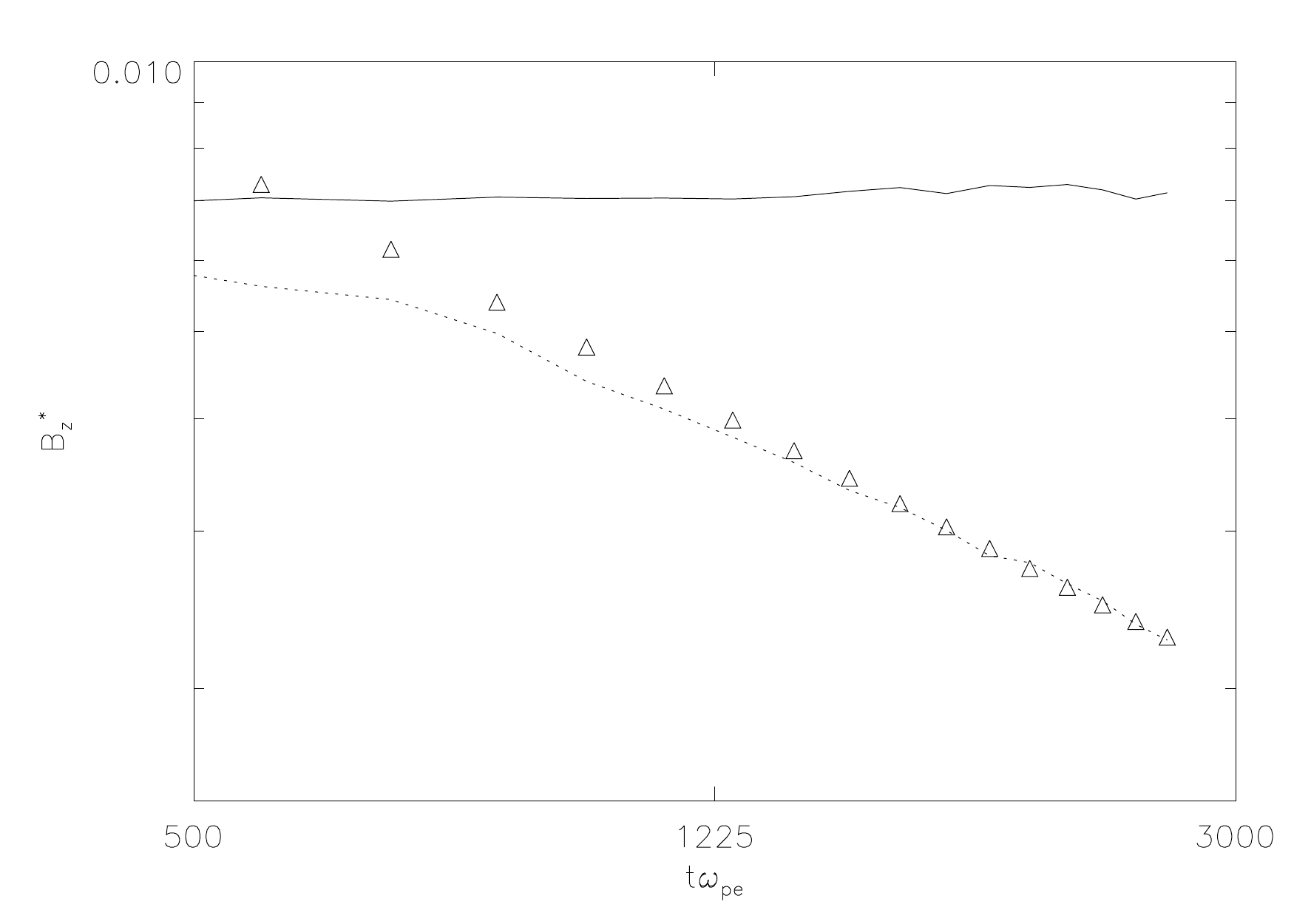}}
\caption{As in Fig.~\ref{fig4} but for the run 2. Here the numerical fit produces the following scaling
law $B_z^*\propto t^{-0.75}$.}
\label{fig10}
\end{figure}
Fig.~\ref{fig10} shows the pulse
amplitude dynamics as in Fig.~\ref{fig4} but for the run 2. In this 
case the numerical fit produces the scaling
law  of $B_z^*\propto t^{-0.75}$. This exponent is note quite $-1$ but
reasonably close. The discrepancy could be due to the fact the the excited
pulse is not quite an eigen-mode of the system and also the non-linearity
could play role as the pulse amplitude is $20\%$, while the $B_z^*\propto t^{-1}$
is according to the linear KdV equation.

\begin{figure*}[htbp]    
\centerline{\includegraphics[width=0.75\textwidth]{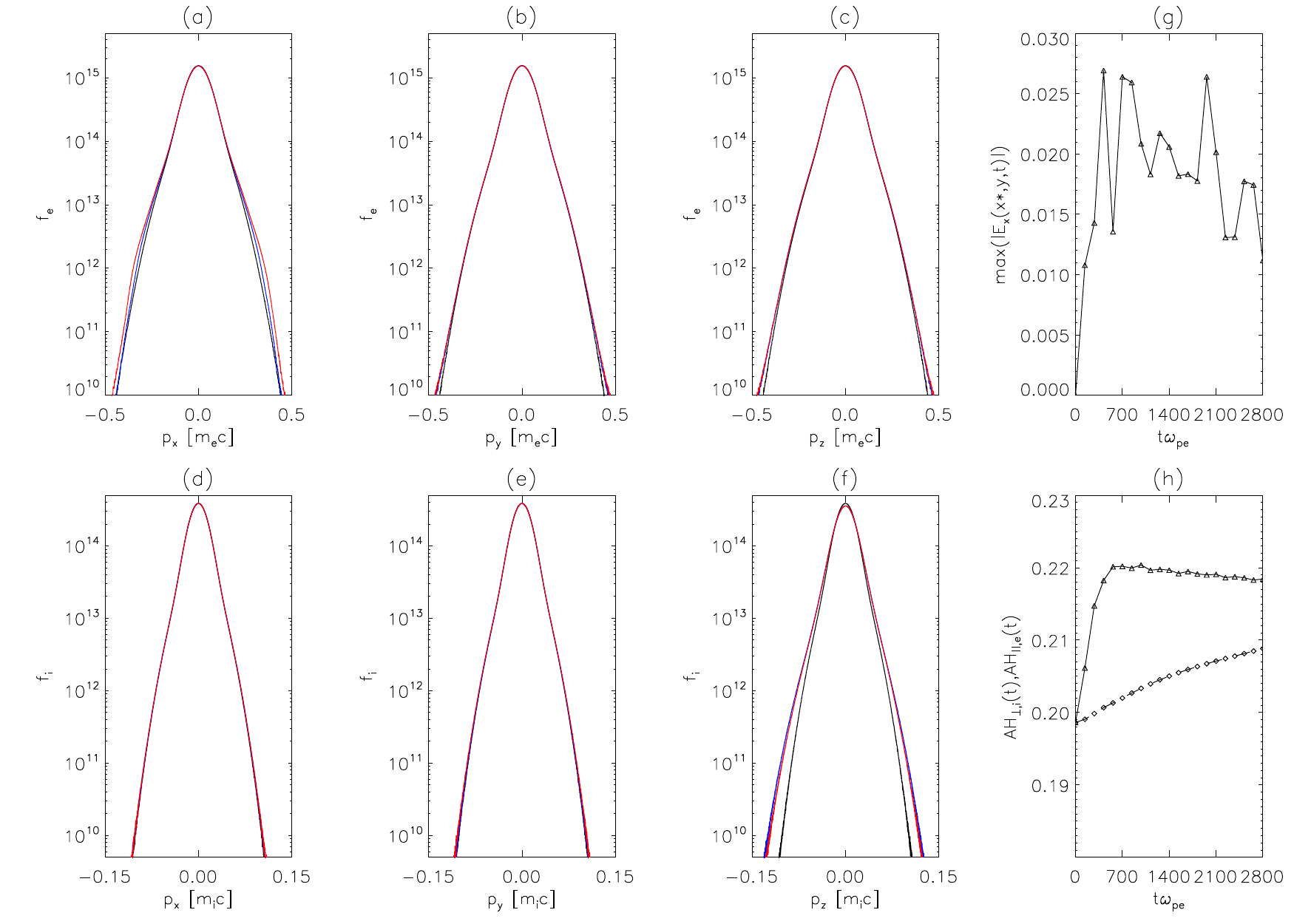}}
\caption{As in Fig.~\ref{fig5} but for the run 2.}
\label{fig11}
\end{figure*}
There are many similarities of Fig.~\ref{fig11} to Fig.~\ref{fig5},
except that widening of the distribution function for ions in 
$p_z$ direction (panel f) is now symmetric.

\begin{figure}[htbp]    
\centerline{\includegraphics[width=0.5\textwidth]{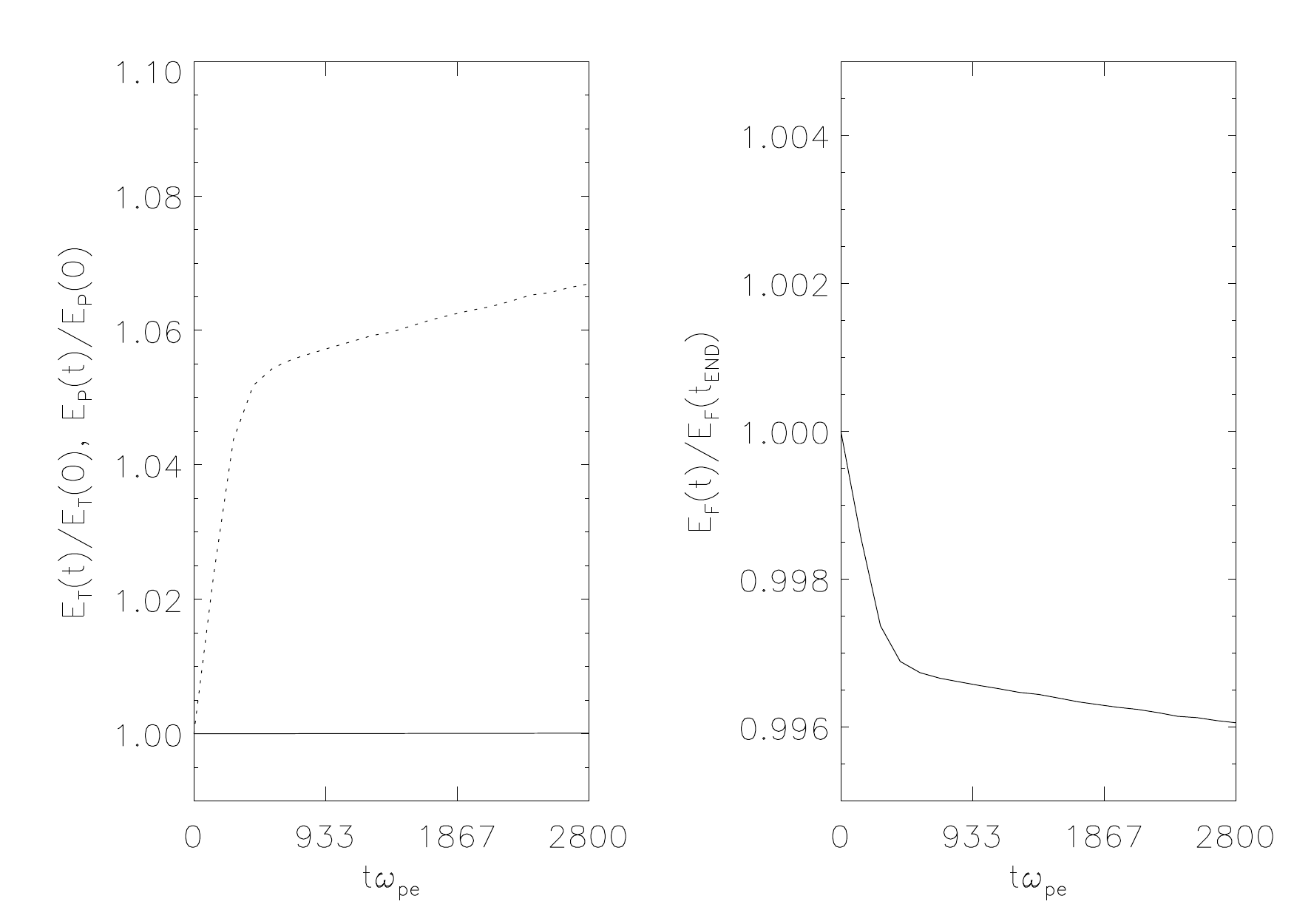}}
\caption{As in Fig.~\ref{fig6} but for the run 2.}
\label{fig12}
\end{figure}

In Fig.~\ref{fig12}, compared to Fig.~\ref{fig6}, the following
modifications can be observed: because there is no continuous
energy input into the system, and  we rather deal with initial value
problem according to equation~\ref{eq4}, there is monotonous
increase in particle energy, while electromagnetic energy
energy does not increase and it only decreases.
The total energy line is nearly flat this due to the fact
that the energy error is 0.0009\%.

\begin{figure*}[htbp]    
\centerline{\includegraphics[width=0.75\textwidth]{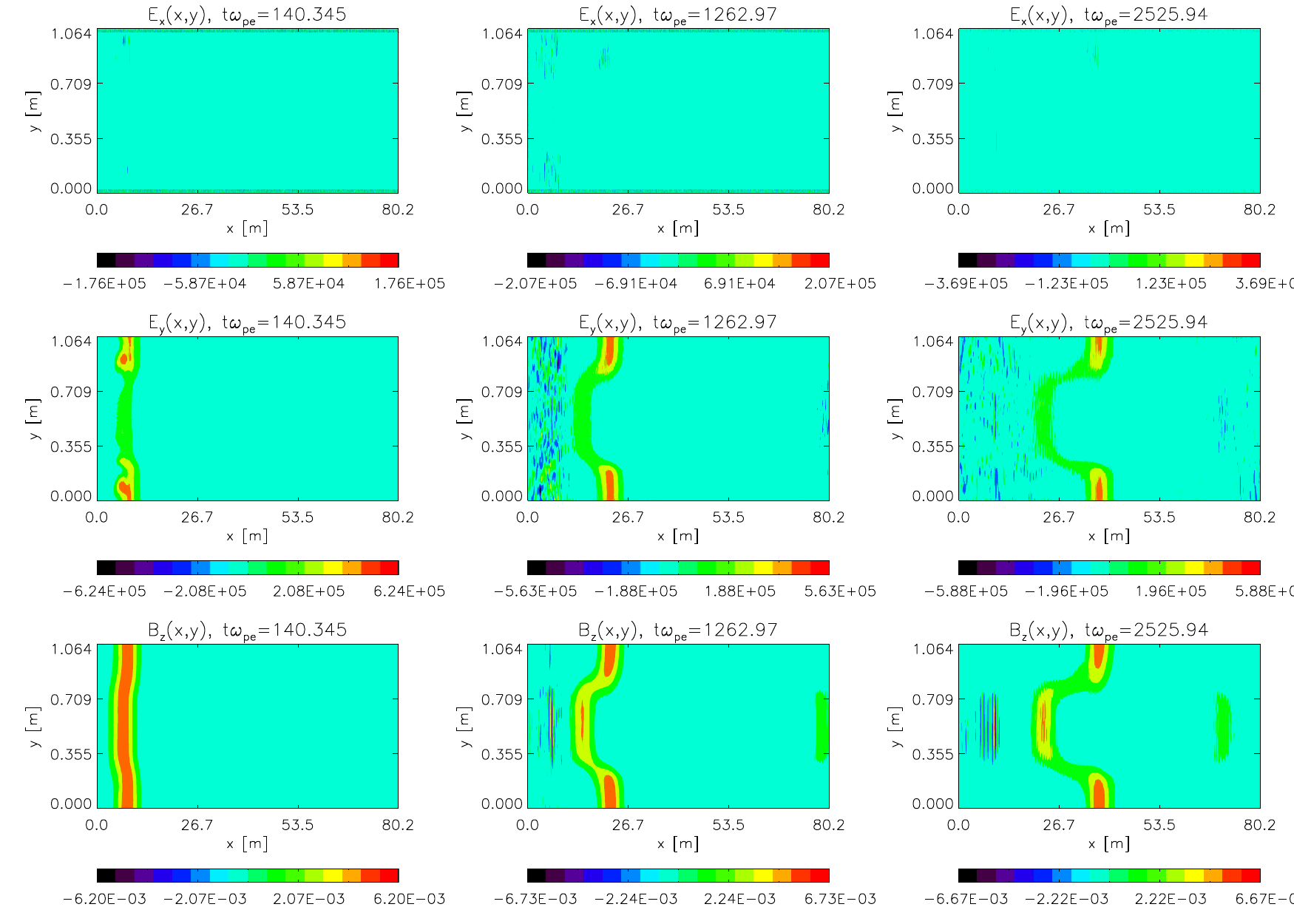}}
\caption{As in Fig.~\ref{fig3} but for the run 3.}
\label{fig13}
\end{figure*}

Run 3 is our best attempt to launch a single Gaussian pulse that
propagates in the positive x-direction. This is achieved by the initial
condition specified above. In Fig.~\ref{fig13} 
we plot time evolution of phase mixed electromagnetic components
$E_y$  and $B_z$ in a similar manner as in Fig.~\ref{fig3}.
We see from these panels that a single, Gaussian pulse is excited moving
to the right (positive x-direction), only a minor backwards propagating
pulse is visible in two bottom right panels near $x=80$ and $x=70$.

\begin{figure}[htbp]    
\centerline{\includegraphics[width=0.5\textwidth]{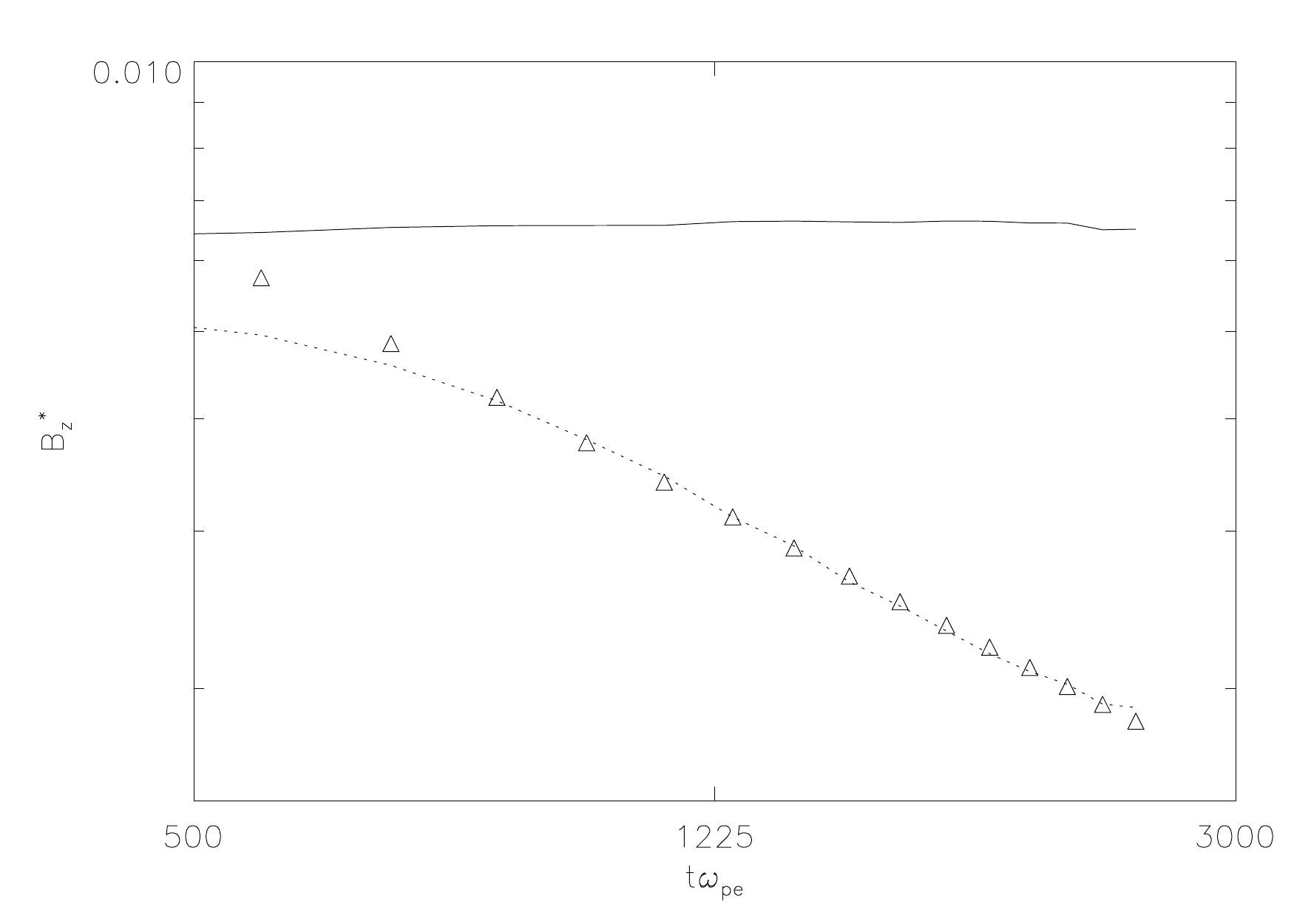}}
\caption{As in Fig.~\ref{fig4} but for the run 3. Here the numerical fit produces the following scaling
law $B_z^*\propto t^{-0.76}$.}
\label{fig14}
\end{figure}
In Fig.~\ref{fig14} we plot the pulse
amplitude dynamics as in Fig.~\ref{fig4} but for the run 3. In this 
case the numerical fit produces the scaling
law  of $B_z^*\propto t^{-0.76}$. Again, this exponent is note quite $-1$ but
tolerably close. 

\begin{figure*}[htbp]    
\centerline{\includegraphics[width=0.75\textwidth]{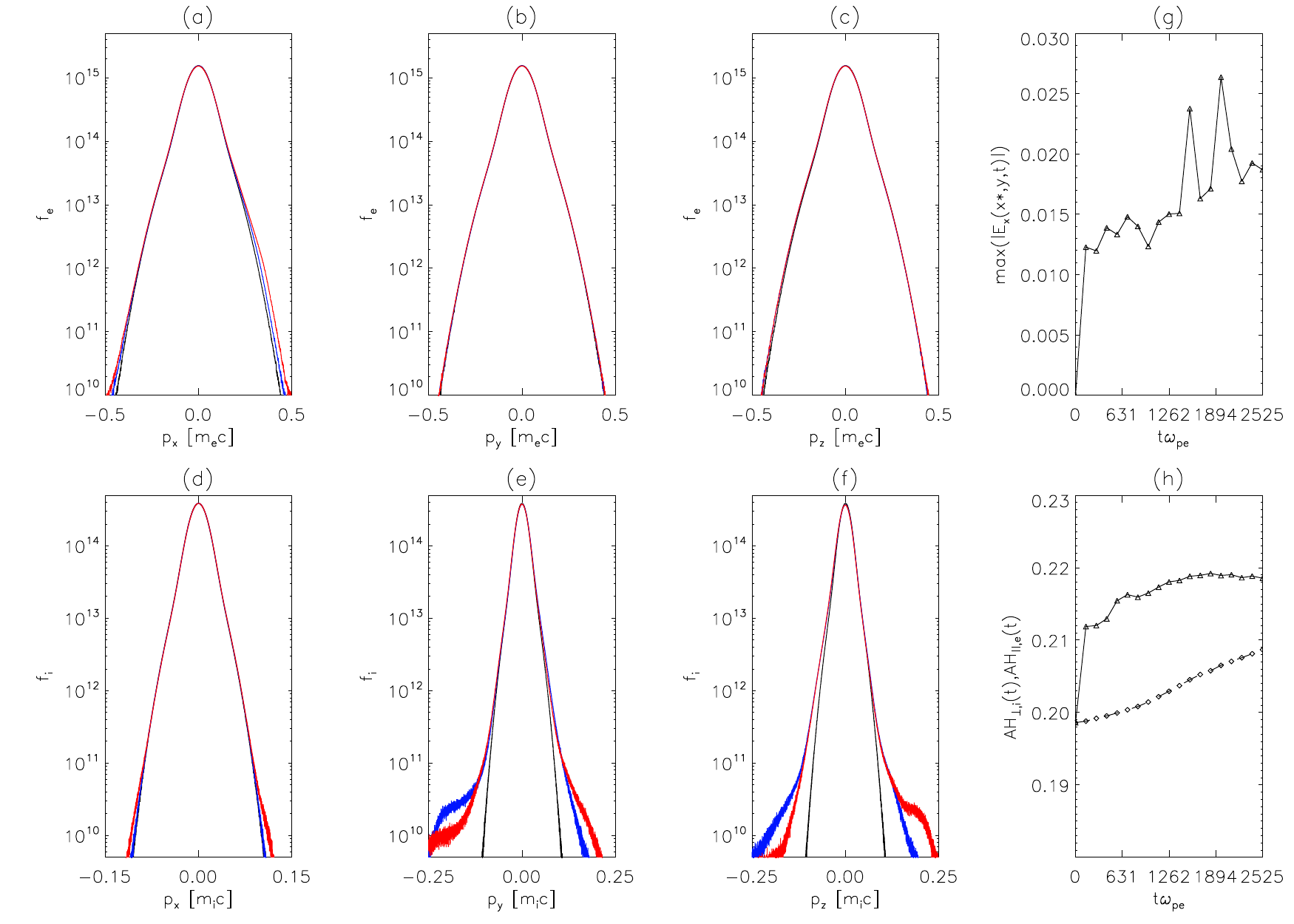}}
\caption{As in Fig.~\ref{fig5} but for the run 3.}
\label{fig15}
\end{figure*}

Fig.~\ref{fig15} panel (a) shows that as the time progresses
the pump in the parallel electron distribution function
develops is single pump corresponding to the positive velocity 
of the Gaussian pulse $V_{phase}/c =0.25$. We explain this by the fact
the in Run 3 only one pulse is present that moves to positive x-direction.
This is an 
interesting result partly because now we see bump only in the positive velocities, contrary to our earlier
works \citet{2005A&A...435.1105T,2008PhPl...15k2902T,2011PhPl...18i2903T,2012PhPl...19h2903T}.
This serves a further proof that the particle acceleration  is via Landau resonance damping.

\begin{figure*}[htbp]    
\centerline{\includegraphics[width=0.75\textwidth]{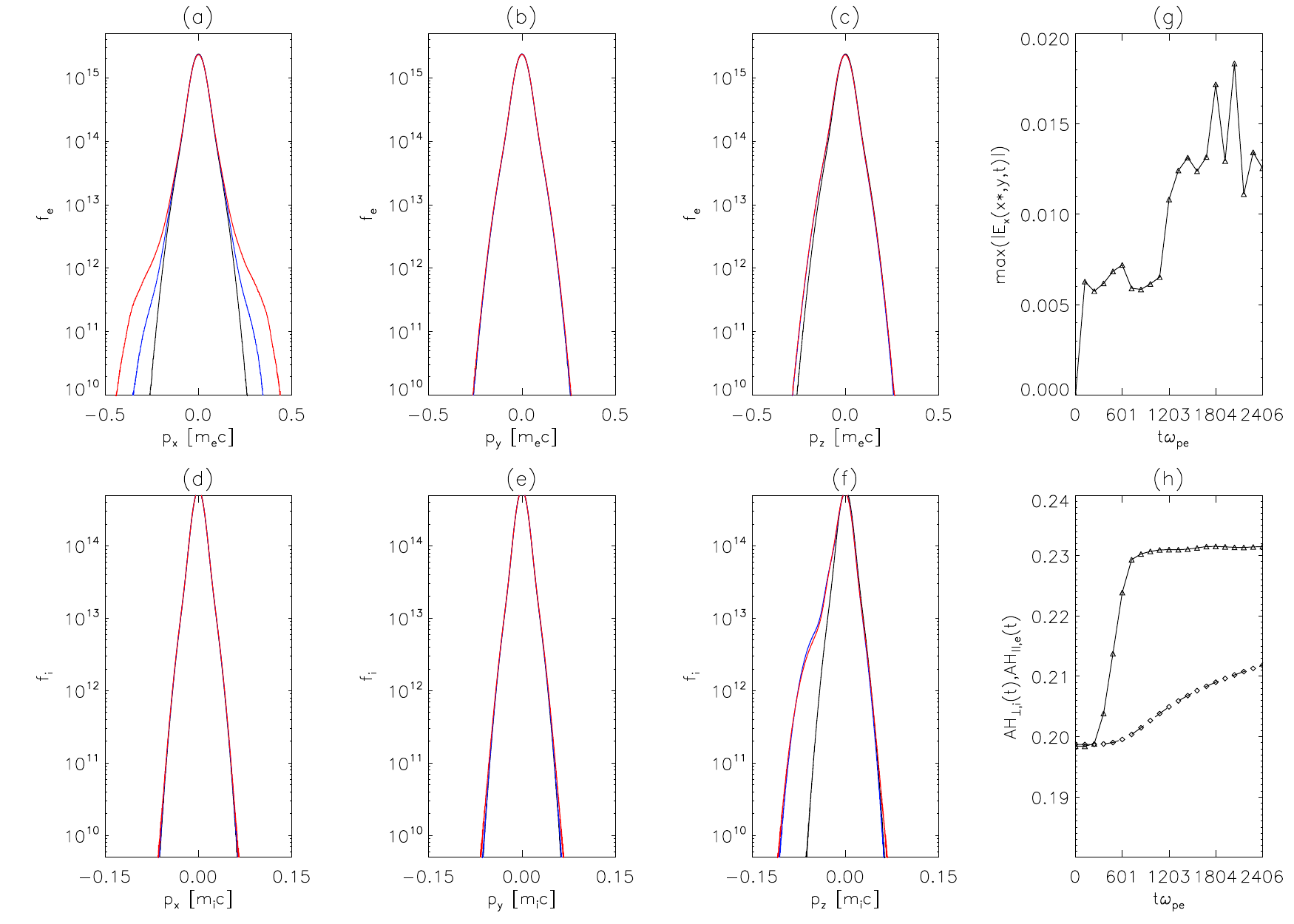}}
\caption{As in Fig.~\ref{fig5} but for the run 1C.}
\label{fig16}
\end{figure*}

For run 3 the behaviour of the energies is similar
to run 1, so no shown here -- particle energy increases on the expense
of the decrease of magnetic energy as the pulse damps.
The total energy line is nearly flat this due to the fact
that the energy error  is 0.02\%. 

In Figure~\ref{fig16} we plot distribution function time evolution for Run1C.
It is worthwhile to note the wider spread of red curve in panel (a)
compared to panel (a) from Figure~\ref{fig5}. This means that in cooler
background plasma with $T=2\times10^6$ (note that all other runs in this work
have three times higher temperature of $T=6\times10^6$).
We gather from panel (h) that 
 $AH_{||,e}$ index  starts from $0.199$ and stops at $0.212$, meaning that $0.013$, i.e.
1.3\% of electrons are accelerated above thermal speeds.
For ions this number is about three times as large ($0.231-0.199\approx 3.2\%$). 

\section{Conclusions}

We have advanced the knowledge of 
dynamics of Alfven waves and associated particle 
acceleration in the inhomogeneous space and solar plasmas
by considering new type of DAW in a form of the Gaussian pulses. 
The latter 
are more appropriate for
solar flares, as the flare is impulsive in its nature.
The  new results can be summarised as following:
(i) Our linear analytical model makes a prediction that the pulse
amplitude decrease is described by the linear KdV equation. 
(ii) The numerical and analytical solution
of the linear KdV equation shows the pulse amplitude damping in time as $t^{-1}$,
which is corroborated by full PIC simulations. 
(iii) We also prove that  the electron acceleration 
is due to collisionless Landau damping. 
However, we would like to stress that the pulse
amplitude decrease (the $t^{-1}$ scaling law) is due to dispersive effects.
In effect, Section 3.1 shows that the phase-mixing leads to the dispersion, described by KdV equation, 
without resorting to the 
damping and wave-particle resonance.
The dynamics of the particle
distribution function in Section 3.2 shows that
 the particle resonance is with the phase-mixed waves which have the 
 phase speed of the DAW pulse. 
(iv)  We show that 
reducing background plasma temperature 
yields more efficient particle acceleration.
(v) When we considered four times more massive ions with $m_i/m_e=64$,
compared to the most runs in this study with $m_i/m_e=16$,
this resulted in four times more efficient
electron acceleration i.e. 4\%. At this stage it is impossible
for us to simulate the realistic mass ratio of $m_i/m_e=1836$ due to the
computational limitations. Thus the jury is still out in the
issue of feasibility of efficient electron acceleration by means
of Gaussian, Alfvenic pulses. 
It should be noted that the issue of scaling
of the generated $E_\parallel$ and hence the particle acceleration
 with the mass ratio has
been investigated before \cite{1367-2630-9-8-262}
for the case of harmonic DAW. 
Ref.\cite{1367-2630-9-8-262} proved that the minimal model 
  required to reproduce previous kinetic 
  results on $E_\parallel$ generation is 
  the two-fluid, cold plasma approximation in the linear regime. 
Ref.\cite{1367-2630-9-8-262} established that 
  amplitude attained by $E_\parallel$ decreases linearly as the inverse of 
  the mass ratio $m_i / m_e$ , 
  i.e. $E_\parallel \propto 1/ m_i$ . This result 
  contradicts the earlier works \cite{angeo-22-2081-2004,JGRA:JGRA14612} 
  in  that the 
  cause of $E_\parallel$ generation is the polarization drift 
  of the driving wave, which scales as $E_\parallel \propto m_i$. 
  Increase in mass ratio does not have any effect on the final parallel 
  (magnetic field aligned) speed attained by electrons. 
  However, parallel ion velocity decreases linearly with the 
  inverse of the mass ratio $m_i / m_e$ , i.e. the parallel velocity 
  ratio of electrons and ions scales directly as $m_i / m_e$. 
  These were interpreted as follows: 
  (i) ion dynamics plays no role in the $E_\parallel$ generation; 
  (ii) decrease in the generated parallel electric field 
  amplitude with the increase of the mass ratio $m_i / m_e$ is 
  caused by the fact that the harmonic driving frequency
  $\omega_d =0.3 \omega_{ci} \propto 1/m_i$
   is decreasing, and hence the electron fluid can effectively 'short-circuit' 
   (recombine with) the slowly oscillating ions, hence producing smaller 
   $E_\parallel$ which also scales exactly as $1/ m_i$.  
Evidently, the same argument does not apply when harmonic
DAW is replaced by the Gaussian pulse, which has no  
"driving frequency" associated with it. Thus, further work is needed
to investigate the scaling of the particle acceleration 
with the mass ratio.

Yet another issue that needs to be mentioned is the fact that
flaring solar coronal plasma has plasma beta possibly close to 
unity ($\beta=p_{gas} / (B^2/2\mu_0) \approx 1 \gg m_e/m_i$) 
\cite{2005psci.book.....A}. Whereas in our work 
most numerical runs are done for 
 $\beta=0.02 < m_e/m_i=1/16=0.0625$ and one run with
 Thus $\beta=0.02 > m_e/m_i=1/64=0.0156$.
We would like to remark that plasma conditions
where the flare occurs and DAW propagate
can be quite different. Once DAW leave the flare
cite with $\beta \approx 1$ after their excitation,
when rushing towards the footpoints, as sketched 
in Figure \ref{fig1}, they will be moving through
plasma with $\beta \ll 1$. As already stated above,
a separate study needs to be conducted how 
particle acceleration efficiency scales with
the mass ratio, i.e. different relations between $\beta$
and $m_e/m_i$.

\begin{acknowledgments}

David Tsiklauri would like to cordially thank 
(i) Japan's 
National Institute of Information and Communications Technology (NICT)
for guest researcher award during 1--31 May 2016 visit to
Toyama University, Toyama, Japan and (ii)
Japanese colleagues Yasuhiro Nariyuki, Takayuki Haruki, Jun-Ichi Sakai
for their kind hospitality and useful, stimulating
discussions during the visit.
This research utilized Queen Mary University of London's (QMUL) 
MidPlus computational facilities,       
supported by QMUL Research-IT and funded by UK EPSRC grant EP/K000128/1.
EPOCH code development work 
was in part funded by the UK EPSRC grants 
EP/G054950/1, EP/G056803/1, EP/G055165/1 and EP/ M022463/1, to which author
has no connection.
\end{acknowledgments}


\end{document}